%% file: MeasGravitationalEffectsOfPulsedBeams.tex
\documentclass[aps,pra,preprint,amsmath,amssymb,superscriptaddress,nofootinbib]{revtex4}
\usepackage{graphicx,color}
\usepackage{dcolumn}
\usepackage{bm}
\usepackage{braket}
\usepackage{ifsym}
\usepackage[english]{babel}
\begin{document}
\title{Perspectives of measuring gravitational effects of laser
	light and particle beams}
\author{Felix Spengler}
\email{Felix-Maximilian.Spengler@Uni-Tuebingen.de}
\affiliation{Eberhard-Karls-Universit\"at T\"ubingen, Institut f\"ur Theoretische Physik, 72076 T\"ubingen, Germany}
\author{Dennis R\"atzel}
\email{Dennis.Raetzel@ZARM.Uni-Bremen.de}
\affiliation{Humboldt Universit\"at zu Berlin, Institut für Physik, Newtonstra{\ss}e 15, 12489 Berlin, Germany}
\affiliation{ZARM, Universität Bremen, Am Fallturm 2, 28359 Bremen, Germany}
\author{Daniel Braun}
\email{Daniel.Braun@Uni-Tuebingen.de}
\affiliation{Eberhard-Karls-Universit\"at T\"ubingen, Institut f\"ur Theoretische Physik, 72076 T\"ubingen, Germany}
\centerline{May 24, 2022}
\begin{abstract}
	We study possibilities of creation and detection of  oscillating gravitational fields from lab-scale high energy, relativistic sources. The sources considered are high energy laser beams in an optical cavity and the ultra-relativistic proton bunches circulating in the beam of the Large Hadron Collider (LHC) at CERN. These sources allow for signal frequencies much higher and far narrower in bandwidth than what most celestial sources produce.  In addition, by modulating the beams, one can adjust the source frequency over a very broad range, from Hz to GHz.
	The gravitational field of these sources and responses of a variety of detectors are analyzed. 
        
         We optimize a
mechanical oscillator such as a pendulum or torsion balance as detector
and find parameter regimes such that -- combined with the planned high-luminosity upgrade of the LHC as a source --
a signal-to-noise ratio substantially larger than 1 should be achievable at least in principle, neglecting all sources of technical noise.   
    This opens new perspectives of studying general relativistic effects and possibly quantum-gravitational effects with ultra-relativistic, well-controlled terrestrial sources. 
\end{abstract}
\maketitle
\section{Introduction}

With the successful measurement of gravitational waves through the LIGO/Virgo collaboration,
the measurement of gravitational signals from relativistic sources has
gained a lot of interest as it is believed to lead
to new insights about gravity, in particular, constraints on modifications 
of general relativity and potential effects of quantum gravity 
\cite{berti2018extreme,berti2018extreme,jimenez2018born,PhysRevD.97.061501,barcelo2017gravitational,PhysRevLett.120.081101}.
However, such experiments are limited
to detection since the experimenter has no access to the cosmic sources of the signal.

Starting already in the 1970s, proposals were formulated for 
constructing terrestrial relativistic sources and detectors of their
gravitational signals. 
E.g.~in
\cite{grishchuk_emission_1974,grishchuk_excitation_1975,grishchuk_gravitational_1977}
a cylindrical microwave resonator was proposed as source of a standing 
gravitational wave and a second 
concentric cylinder as detector
based on photon creation in one of its modes. But it was clear that
with the existing technology at the time it was not realistic to
create a sufficiently strong source whose radiation could be
detected. In recent years there has
been renewed interest in the creation and detection of 
gravitational waves in the lab \cite{portilla_generation_2001,ballantini_detector_2003,grishchuk_electromagnetic_2003,rudenko_very_2004,baker_utilization_2012,kolosnitsyn_gravitational_2015,fuzfa_electromagnetic_2018,chen2021srgw,jowett2021srgw}. \\ 

The gravitational field of electromagnetic radiation has been studied early on 
\cite{tolman_1931,bonnor_1969, aichelburg_1971}.
It gives rise to a range of interesting effects, from an
attraction that decays with the inverse of the distance instead
of the inverse square \cite{tolman_1931,bonnor_1969,aichelburg_1971,ratzel_gravitational_2016}
to frame dragging \cite{mallett_2000,strohaber_2013}
and other gravitomagnetic effects \cite{cox_2007,scully_1979,ji_1998_gravitational,ji_2006_gravitational,schneiter_gravitational_2018,schneiter_corrigendum:_2019,schneiter_rotation_2019}
. Their  detection has been found to be extremely challenging, see
e.g.~\cite{scully_1979,ji_1998_gravitational,ji_2006_gravitational,ji_2007_gravitational,strohaber_2013,ratzel_gravitational_2016,schneiter_gravitational_2018,schneiter_corrigendum:_2019,schneiter_rotation_2019}.
The phenomenology of the gravitational field of relativistically moving matter is similar to that of light.
It can be calculated by Lorentz boosting spacetimes of sources at rest.
The result 
approaches
the gravitational field of massless particles in the ultra-relativistic limit \cite{aichelburg_1971,balasin_1996_boosting,lousto_1990,barrabes_2003,schneiter_gravitational_2018,schneiter_corrigendum:_2019}.

As technology has substantially progressed since some of the cited
works have been published, both on the side of sources in the form of 
high-power lasers
\cite{nakamura_diagnostics_2017,lawrence_livermore_national_laboratory_national_2016,georg_korn_ed_eli_2011}
and particle accelerators \cite{LHC,noauthor_beam_2013},
and in the metrology of extremely weak forces \cite{westphal_measurement_2020,PhysRevA.101.011802,schmole_micromechanical_2016,kapner_tests_2007,hoyle_submillimeter_2004,schreppler_optically_2014},
it is worthwhile to reassess the possibility to detect 
the gravitational effects of light and of ultra-relativistic particle beams. 
Indeed, progress in this direction would enable the test of
general relativity (GR) in a new, ultra-relativistic regime (in the
sense of special relativity), with an energy-momentum tensor as the source
term in Einstein's equations very different from the one that can be
achieved with non-relativistic masses and purely Newtonian gravity,
namely with a large off-diagonal component in Cartesian coordinates.

In this article 
we focus on the acceleration of non-relativistic sensor systems due to 
the gravitational field of light beams and ultra-relativistic particle
beams such as the ones produced at the LHC. 
We add several new aspects that improve the
outlook for experimental observation.  
Most importantly, we consider trapping of laser light in a cavity, 
through which the circulating power can be drastically enhanced. Secondly, we consider
modulation of the gravitational sources with an adjustable frequency in order to match them to the optimal sensitivity of existing detectors.  
Several approaches are investigated to that end.  The simplest one consists in having laser pulses oscillate to and fro in a cavity, 
such that the length of the cavity determines the oscillation frequency of the gravitational signal.  
We also 
examine the possibility of slowly (kHz frequency) modulating the power with which the cavity is pumped using a continuous wave (cw) or pulsed laser. With the pump power, the power circulating in the cavity is modulated, and thus, also the strength of the gravitational field. 
Thirdly, we extend
the analysis to ultra-relativistic particle beams such as available at
the LHC.
And finally, we examine several possible
sensors for their suitability for measuring the created gravitational
fields. 

Our work is also motivated by current developments
towards measuring gravitational effects of sources in a quantum
mechanical superposition
as a possible experimental road to understanding quantum gravitational effects \cite{pikovski_probing_2012,belenchia_quantum_2018,westphal_measurement_2020,schmole_micromechanical_2016}.  
Creating quantum superpositions of sufficiently large masses is
challenging, and it is therefore worthwhile  
to think about other sources that can be superposed quantum
mechanically.   
We discuss perspectives in this direction for the gravitational
sources studied in this paper in Sec.\ref{Sec.discus}.

\section{Potential sources and their gravitational field}
\subsection{Laser pulses oscillating in a cavity}
\begin{figure}[t!]
	\includegraphics[width=\textwidth]{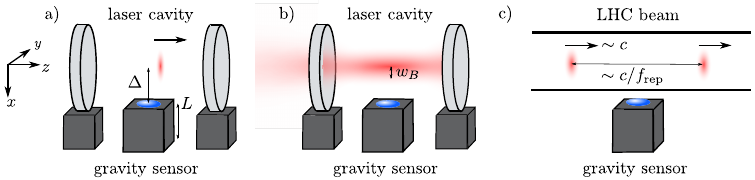}
	\caption{
            a) Laser pulse oscillating to and fro in a cavity.
            b) cw laser focused to a
            narrow waist inside a cavity.  
            Its intensity is
            modulated to create a gravitational field oscillating at
            kHz frequency. 
            c) Ultrarelativistic particle bunches in an accelerator ring such as the LHC create a gravitational field very similar to that of laser pulses. 
            In the vicinity of the waist of the laser beam or close to the beamline, a detector picks up resonant mechanical deformations due to the oscillating gravitational forces.}\label{fig:sourcecavity}
\end{figure}

To create a strong, high frequency gravitational field, a source of
high power and intensity is required. Modern femtosecond laser pulses
can reach up to a Petawatt in pulse power. One such laser pulse
oscillating in a cavity, as illustrated in fig.~\ref{fig:sourcecavity},
is  a source of short bursts of high energy  
oscillating to and fro at high frequency.
The perturbation to the metric and
the resulting Riemann curvature tensor can be calculated within the
theory of linearized gravity, as is done in appendix \ref{appendix:h_reflected_pulse} and \cite{schneiter_gravitational_2018,schneiter_corrigendum:_2019}. 

For a continuous-wave ({\em cw})-laser
with power 
$P$ 
and circular polarization, the curvature component relevant to a
non-relativistic sensor based on a mechanical resonator with axis perpendicular to the beam line of the laser is,  for an observer in the $x$-$z$-plane (i.e. $y=0$) and in the approximation of a vanishing opening angle,
given by 
\begin{equation}
\label{eq:R0x0x}
R_{0x0x}\simeq -4G P/(c^5 \rho^2)  
\end{equation}
with $G$ the gravitational constant, $c$ the speed of light in vacuum, $\rho^2=x^2+y^2=x^2$, and $ x_0 = ct $. 

Laser {\em pulses} were considered earlier in \cite{tolman_1931,ratzel_gravitational_2016} in the
approximation of an infinitely thin light pencil of length $L$. 
A further exploration in appendix \ref{appendix:h_reflected_pulse} for the simplified case of box shaped pulses oscillating to and fro corroborates the result that close to the beam ($ \rho \ll |z|,|z-D| $, where $ z=0 $ and $ z=D $ are the positions of the two mirrors) eq. \eqref{eq:R0x0x} gives the correct
result, limited, however, to a finite duration
of the order of the length of the pulse (see Fig. 8 in 
\cite{ratzel_gravitational_2016}) but on the other hand with the cw-power $P$ replaced
by the power of the effective pulse in the cavity $P_{\rm p}^\mathrm{cav}$ (see eq.\eqref{eq:R_oscill_pulse}).
The curvature results in a tidal force between two infinitesimally separated points next to the beamline.
However, there is no gravitational wave generated as this type of source is not quadrupolar in nature. 
Rather one can detect the gravitational near-field. 

The average power at a given cross-section of the beam inside the cavity is $ {P}_{\rm cav}^\mathrm{avg} =
\frac{2\tau_{\rm p}}{\tau_{\rm rt}} P_{\rm p}^{\rm cav} $, where $ \tau_{\rm p} $ is the length of the pulse, and
$ \tau_{\rm rt} = \frac{2L_{\rm cav}}{c} $ is the round trip time in the
cavity. As the power enters linearly into the gravitational potential,
acceleration, and curvature, the considered gravitational effects will be
proportional to the average power 
in the cavity.

\label{sec:pulse_cavity}
A pump laser emitting very short pulses has a broad spectrum in 
the frequency domain. 
Coupling these pulses into a cavity of high finesse $ F \approx \frac{\pi (R_1R_2)^{\frac{1}{4}}}{1-\sqrt{R_1R_2}} $
 \cite{ismail2016fabry} leads to an electric field strength inside the cavity
\begin{equation}\label{eq:cavitytransferfunction}
	\tilde{E}_\mathrm{cav}(\omega) = \tilde{G}_{\rm cav} (\omega) \tilde{E}_{\rm p} (\omega),\quad \mathrm{where} \quad \tilde{G}_\mathrm{cav}(\omega) = \frac{\sqrt{T_{1}}}{1-\sqrt{R_1R_2} \exp (-i\omega \tau_{\rm rt})}
\end{equation}
is the field transfer function, with the intensity transmissivity
$ T_{1} $ of the mirror struck by the pump beam, and the intensity
reflectivities of the two mirrors $ R_{1/2} $, where
$T_1=1-R_1$. An explicit calculation for the case of rectangular pulses can be found in appendix \ref{app:c}, which is based on \cite{cesini_1977}.
If the pulses are very short $ \tau_{\rm p} \ll \tau_{\rm rt} $ and
far apart $ 1/f_{\rm rep} \gg \tau_{\rm L}  $, where $ \tau_{\rm L} \approx
\frac{2 F}{\pi} \tau_{\rm rt} $ is the $ 1/e $ energy decay time and $ f_{\rm rep} $ is the repetition rate of the pump laser,  the
pulse enters the cavity at an intensity $ T_1 I_{\rm p} $, where the circulating power is enhanced by a factor $ \frac{2 F}{\pi} $, independent of the cavity length. 
 Without any further modification the factors $T_1$ and $ \frac{2
   F}{\pi} $ at best  cancel up to a factor of 4 (assuming $R_1,R_2\simeq 1$), leaving little to be gained (see App.\ref{app:c}, eq. \eqref{eq:cavityenhancement}). 
 The ways one could imagine improving upon this all involve changes to the mirror that couples the pump laser pulses to the cavity:
 \begin{itemize}
 \item An input coupler is a mirror which is significantly less
   reflective than what could be achieved with the best available
   mirrors. Combined with techniques such as impedance matching, it increases the power deposited into the cavity while also slightly reducing the cavity finesse \cite{pupeza2012power}.  
   For example, in \cite{carstens_2014} input couplers are employed to realize enhancement cavities with kilowatt-average-power femtosecond pulses, increasing the average power circulating in the cavity to $ 670 $ kW, $ 10^3 $ times the $ 420 $ W average power of the pump laser. 
   Using larger laser spots on the mirrors of the cavity should allow for even stronger pump lasers to be used. With stronger pump lasers, such as the BAT laser in \cite{10.1117/12.2525380} with an average pump power of $ P_{\rm pump} = 300 $ kW, an average power within the cavity in the 100 MW range seems plausible.
 	
 \item A switchable mirror would allow for the full pump beam power to enter the cavity, which means the average cavity power is expected to be the pump laser power enhanced by a factor $ \frac{2F}{\pi} $. Depending on the cavity's length and the pump laser's repetition rate, the mirror has to be moved on a timescale of $ 10^{-9} $ s to $ 10^{-3} $ s, the slower end of which seems realistic. A mirror mounted on  some mechanics might reduce the precision of its positioning and hence the cavity's finesse. Nonetheless, with a high finesse cavity ($ \frac{2 F}{\pi} \sim 10^5 $) and high-average-power pump lasers ($ P_{\rm pump} \approx 300$ kW \cite{10.1117/12.2525380}
 	) an average power $ > 20 $ GW in the cavity would be achievable.
 \end{itemize}

 One limitation when scaling to higher powers is damage to the mirrors. In \cite{schwartz_near-concentric_2017} the cw intensity threshold was determined to be at around $ 100\, {\rm MW}/{\rm cm^2} $ before thermal damage sets in. For sub-picosecond pulses the intensity threshold can be exceeded by at least an order of magnitude, as it is done in \cite{carstens_2014}, as long as the average intensity on the mirrors does not exceed the thermal threshold. For 20 GW (100 MW) cavity power this 
 needs a spot diameter on the mirrors of at least 16 cm (1.1 cm). For the input coupler the limitations are even stricter than for the end mirror as the power passes through the input coupler and creates more heating than when reflected at the surface of the reflecting mirror \cite{carstens_2014}. Large spot sizes require long cavities, as otherwise the mode in the cavity has a large opening angle and prevents positioning the sensor very close to the beam. For the cited spot-sizes of order 1-10\,cm, a cavity length $ L_{\rm cav} \gtrsim 1 $\,m suffices. 
In this work, the increase in power is accounted for by increasing the pulse
duration by defining an effective pulse length 
\begin{equation}\label{eq:effectivepulselength}
	T^{\rm cav}_{\rm  p}=T^{\rm pump}_{\rm p} f_{\rm rep} \tau_{\rm rt}\frac{2F}{\pi} =T^{\rm pump}_{\rm p} f_{\rm rep} \tau_{\rm L}.
\end{equation}
For the BAT laser from \cite{10.1117/12.2525380}, the repetition rate is $ f_{\rm rep} = 10 $ kHz and the pulse duration of the pump laser is $ T^{\rm pump}_{\rm p} = 100 $ fs. We further assume $ F = 10^5 $ and the signal to be at resonance with the sensor frequency $ \tau_{\rm rt} = 2 \frac{2 \pi}{\omega_0} $, see
e.g.~Table~\ref{tab:powers}.  This is consistent with the image of
creating a ``train'' of pulses (one could also imagine pulse stacking,
i.e. increasing the pulse power instead).  
Laser pulses with far higher pulse powers exist.   The National Ignition Facility achieves $ 5\cdot 10^{12} $\,W peak power 
\cite{lawrence_livermore_national_laboratory_national_2016} but is not as suitable for our purposes due to its low repetition rates. Peak powers of up to $10\cdot 10^{15}$\,W at repetition rates of up to 10\,Hz exist \cite{georg_korn_ed_eli_2011} and others with peak powers  on the order of $100\cdot 10^{15}$\,W are planned \cite{cartlidge_physicists_2018}, but will need to achieve higher average intensities and repetition rates in order to lead to measurable gravitational effects. 

\subsection{Modulated cw-pumping}\label{sec:cw_cavity}
Instead of creating a periodic signal by 
having laser pulses oscillate in a cavity, one could also consider using a cw laser. To create a periodic signal, one can pump the cavity for part of the period and allow for the intensity inside the cavity to decay before switching the pump beam back on for the next period, thus creating a modulated signal with modulation period $ \tau_{\rm mod} $.
Depending on $ \tau_{\rm L}$, the energy within the cavity as a
function of time looks more like a   periodic sequence of effective pulses that have the form of rectangles --- in the case of $ \tau_{\rm L} \ll \tau_{\rm mod} $ --- or like a series of shark fins for $ \tau_{\rm L} \sim \tau_{\rm mod} $ (see appendix \ref{app:c}).
We call $P_{\rm p}^{\rm cav}$ the maximum power of the effective pulse in the cavity.

Using a cw pump laser, the coupling to the cavity is no longer detrimental as for $ \Delta \omega_{\rm FWHM} \sim 1/\tau_{\rm L}  > \Delta\omega_{\rm pump} $, where $ \Delta\omega_{\rm pump} $ is the line width of the pump beam, the pump beam couples almost fully to the cavity.
The Newtonian gravitational potential for a thin 
light pencil in the form of a standing light wave 
in the cavity is (see \cite{tolman_1931} and appendix \ref{appendix:h_reflected_pulse}) $ \varPhi = \frac{4GP(t)}{c^3} \ln \rho $, 
where $ \rho $ is the distance from the beam line and $ P(t) $ is the power passing through the cross section with the detector, i.e. $ P(t) = P_{\rm cav} (t) $ in this case. 
For the 
slowly moving detectors envisaged here (speeds $v\ll c$), all equations of motion are the same for the source consisting of the standing wave or the propagating one. 

For long modulation periods $ \tau_{\rm mod} \gg \tau_{\rm L} $, the maximum power of the effective pulse in the cavity is $ P_{\rm p}^{\rm cav} = \frac{2F}{\pi} P_{\rm pump}$, for approximately half the modulation period.
Commercially available cw laser systems reach continuous powers of 500 kW in multi-mode operation and up to 100 kW in single-mode operation (see \footnote{
    A single-mode has the advantage that one can focus it down to a spot size comparable to the wave length, i.e. one could get, at least in principle, much closer to the beam (oder 1 $\mu$m instead of ca. 100 $\mu$m. Thus, while loosing a factor 25 in power one gains a factor 100 in distance, i.e. there is an overall improvement by a factor 4 over the multi-mode case, if such small distances from the beam can indeed be realized
}, and e.g.~\cite{ipg}).  Combining this with a high finesse cavity $ F\sim 10^6 $ leads to an average circulating power in the cavity of
\begin{equation}\label{eq:cw_avg_power}
	{P}_{\rm cav}^\mathrm{avg} = \frac{1}{2} \frac{2F}{\pi} P_{\rm pump} \sim 100 \,{\rm GW}.
\end{equation}
The average power in the cavity can at most be a fraction $ < \left(1-e^{-\tau_{\rm mod}/(2\tau_{\rm L})}\right) $ of the maximum power $ \frac{2F}{\pi}  P_{\rm pump} $. 
For slowly decaying cavities, where $ \tau_{\rm mod} \gtrsim \tau_{\rm L}$, techniques such as $Q$-switching or switchable mirrors are necessary to adequately modulate the amplitude \footnote{
    Shorter cavities lead to lower $ \tau_{\rm L} $ at the same finesse and without decreasing the average power. The same considerations for the spot size and length of cavity as mentioned for the laser pulses apply also in the cw case for positioning the detector sufficiently close to the beam waist and neglecting higher order effects in the opening angle \cite{schneiter_gravitational_2018}. For a 1\,m-long high-finesse cavity ($F \sim 10^6$) the decay time $ \tau_{\rm L} $ is in the low millisecond range, which is too slow for some of the proposed detector setups. This could be circumvented by implementing techniques such as Q-switching, with which  the decay of energy within the cavity can be accelerated, and the aforementioned switchable mirrors. Also, the energy buildup can be modulated  to a certain degree by pumping. For  short modulation periods, $ \tau_{\rm mod} \approx \tau_{\rm L} $, the cavity is never fully pumped.
}.

\subsection{The Large Hadron Collider (LHC)}\label{Sec.LHC1}
Instead of 
laser light, one can also investigate ultra relativistic particle beams consisting of high-energy bunches, such as the one at LHC, as gravitational sources.
A  particle beam in the relativistic limit is, from a gravitational perspective, the same as a laser beam: for example, the rest mass of the protons $m\simeq 938\,$MeV/$c^2$ makes a negligible contribution to their energy for achievable particle energies of about 6.5\,TeV and both charge and spin are irrelevant \cite{lousto_1990,balasin_1996_boosting,barrabes_2003}.  To very good approximation, the energy-momentum relationship is then $E=c p$ where $p$ is the momentum of the protons, just as for photons. 
In the ring of the LHC there are $2808$ bunches of protons at maximum capacity, each bunch 
with a total energy of $ \sim 10^5 {~ \rm J}$.
One bunch is approximately $ 30 $ cm long, contains $ 1.15\cdot 10^{11} $ protons, and can be squeezed down to a transverse diameter of $ \sim 16 {~ \rm \mu m}$ (see \cite{LHC}).
To excite a resonator at its eigenfrequency $\nu_0=\omega_0/(2\pi)$, 
the bunches have to pass by the detector with rate $\nu_0$, or the beam must be modulated with frequency $\nu_0$. The $ 2808 $ bunches spread over a ring of 
26,659\,m 
length moving at speed close to $c$
entail a rate of $31.2$\,MHz. 
A single bunch going around the ring passes with a frequency of $  11  $ kHz.  To achieve lower frequencies one 
could, for example, periodically modulate the beam position. This would result in a scheme similar to that of the cw laser cavity, where the LHC beam is active for half the sensor's oscillation period $ \tau_{\rm p} = \frac{1}{2} \frac{2 \pi }{\omega_0} $ with an effective pulse power
$ P_{\rm p}^{\rm cav} = P^{\rm LHC}=2 P^{\rm avg}_{\rm cav}$, where $P^{\rm LHC} $ is the nominal average power of the LHC.
The pulse power of the LHC beam is orders of magnitude smaller than that of extreme-power laser pulses, but the proton bunches are much longer ($\sim 1 $ ns) than the laser pulses.
This results in a higher average power of $ P^{\rm avg}_{\rm cav} \approx 3.8 \cdot 10^{12} $ W, which is orders of magnitude larger than the average power of laser pulses oscillating in a cavity and about 40 times the average power that can be contained in a cavity pumped by 
the cw laser considered above 
(see table \ref{tab:powers}).
Therefore, from the perspective of the strength of the gravitational source, the LHC beam might be preferable.  A potential drawback compared to the laser-based sources is the lack of flexibility in frequency. This can be compensated, however, by considering detectors with tunable resonance frequency.  Besides protons, it is also possible to use heavy nuclei, or partially ionized heavy atoms.  The latter have the advantage that the corresponding beams can be laser-cooled (see the discussion in Sec.\ref{sec.qmgr}). Upgrades of the LHC to use heavy ions are currently considered \cite{krasny_high-luminosity_2020}, and also under development at Brookhaven National Lab \cite{litvinenko_coherent_2017}.
\begin{table}
\begin{tabular}{|c|c|c|c|c|}
    \hline
     & $ P_{\rm p}^{\rm cav} $ & $ T^{\rm cav}_{\rm p} $ & $ P_{\rm cav}^\mathrm{avg} $ & $ w_{\rm B} $ \\
     \hline
     pulses in cavity & $3\cdot 10^{14} $ W & $ 100 $ fs $\cdot 10$ kHz $\frac{8 \cdot 10^5}{\omega_0}\; {}^\dagger $ & $ 2 \cdot 10^{10} $ W & $ < 100 \;\mu $m \\
     \hline
     cw laser+cavity & $ 2 \cdot 10^{11} $ W & $ \frac{\pi}{\omega_0} $ & $ 1 \cdot 10^{11} $ W & $ <100 \mu $m \\
     \hline
     LHC & $ 10^{14} $ W & $ 10^{-9} $ s ${}^\ast$ & $ 3.8 \cdot 10^{12} $ W & 16 $\mu $m\\
     \hline
\end{tabular}
\caption{Comparison of relevant numbers of the LHC beam and the laser-based sources from \ref{sec:pulse_cavity} and \ref{sec:cw_cavity}: 
  $P_{\rm p}^{\rm cav}$ pulse power, $P_\text{avg}$ power averaged over time,
  $w_{\rm B}$ waist of beam. $\omega_0$ is the desired signal frequency,
  assumed in Sec.\ref{sec.Det} to be one of the resonance frequencies
  of the detector.\\
  $ {}^\dagger $ A switchable mirror is assumed for the
  pulses in the cavity. The pulses are assumed to be effectively stacked together to a larger circulating effective pulse, see eq.\eqref{eq:effectivepulselength}. \\
  ${}^\ast$ The effective pulse length $T^{\rm cav}_{\rm
  p}$ for the LHC corresponds to a single proton bunch, but a much slower modulation of the beam on resonance with $\omega_0$ can be envisaged.\\
}\label{tab:powers}\end{table}

\section{Detectors}\label{sec.Det}
We consider three types of detectors, a mechanical rod, a detector
based on superfluid helium-4 coupled parametrically to a
superconducting microwave cavity, and a mechanical harmonic
oscillator, motivated by the monolithic pendulum from
\cite{matsumoto2019demonstration,catano-lopez_high-2020} and the torsion
balance from \cite{westphal_measurement_2020}, with which recently very high levels of sensitivity for gravitational fields have been reached.
The superfluid helium detector and the monolithic pendulum are 
optomechanical detectors close to the quantum limit.  Quantum optomechanical detectors and different configurations have been studied 
in great detail over recent years, both
theoretically and experimentally  \cite{marquardt2009trend,metcalfe2014applications,millen2020optomechanics}. 
They have been considered for high precision sensing \cite{arcizet2006high} 
in particular, force sensing \cite{ranjit2016zeptonewton} and theoretical work has been performed to derive general limits for sensing of oscillating gravitational fields
with such systems \cite{schneiter_2020_optimalest,qvarfort2020optimal}.
We take the mentioned types of detectors as starting points for examining the question what parameter values would need to be achieved such that they become suitable for measuring the gravitational forces considered in this paper. 

\subsection{Mechanical response of a rod} \label{sec:mechanical}
A spatially dependent gravitational acceleration compresses a 1D deformable resonator according to its Young modulus $ Y $.  
The wave equation for the displacement field $ u(x,t) $, describing the relative position of an element of the rod from its equilibrium location $ x $, is given in \cite[p.416]{maggiore2008gravitational} as
\begin{equation}\label{eq:1dwaveeqn}
	\varrho_m\partial^2_t u(x,t) - Y \partial^2_x u(x,t) = -\varrho_m \partial_x \varPhi(x,t),
\end{equation}
where the resonator is extended in the $ x $ direction, orthogonal to the beam, and $ \varrho_m $ is its mass density.
The length contraction due to modification of
space-time is negligible in comparison to the elastic effect considered here, as it comes with an additional factor ${c_{\rm s}^2}/c^2$ \cite{ratzel_frequency_2018}, where $ c_{\rm s} = \sqrt{Y/\varrho_m} $ is the speed of sound in the rod's material.

The displacement field can be expanded into the spatial eigenmodes
$$w_n (x) =  \cos
\left(\left(n+\frac{1}{2}\right)\frac{\pi}{L}(x-\Delta)\right)$$
of the free equation of motion complying with the boundary conditions, i.e. the tip of the resonator distant from the source was chosen to be fixed in place by the support (hence $ w_n (L+\Delta) $ has to
vanish and $  \partial_x w_n\vert_{x=\Delta} $ has to vanish at the other tip), where $ n \in \mathbb{N}_0$, $ \Delta $ is
the distance of the tip of the rod from the source, and $ L $ is the
length of the resonator (see figure \ref{fig:sourcecavity}).  The spatial eigenmodes are orthonormal with respect to the inner product
$\braket{a|b}=(2/L)\int_\Delta^{\Delta+L}a(x)b(x)\,dx$. The total displacement field is then given by $u(x,t)=\sum_{n=0}^\infty
\xi_n(t)w_n(x)$.
The differential equation for the temporal amplitude $ \xi_n(t) $ resulting from the projection of \eqref{eq:1dwaveeqn} 
onto the $ n $th spatial eigenmode is then given by
\begin{equation}
	\ddot{\xi}_n(t) + \frac{\omega_n}{Q} \dot{\xi}_n(t) + \omega_n^2 \xi_n(t) = -\frac{2}{L} \int_{\Delta}^{L+\Delta}{\rm d}x \, w_n(x) \partial_x\varPhi(x,t) ,
\end{equation}
where $ \omega_n = c_{\rm s} \left(n+\frac{1}{2}\right)\frac{\pi}{L} $ is the frequency of the mode and a linear dissipation  term $\gamma_n \partial_t
u(x,t)$ with rate $ \gamma_n = \varrho_m \frac{\omega_n}{Q} $ was added to equation
\eqref{eq:1dwaveeqn} in order to include dissipation from the elastic modes of the
resonator. 

In the case of resonant excitation, the amplitude of the steady state solution in the lowest eigenmode $ \xi_0 (t) = A(\omega_0) \sin (\omega_0 t) $, reached after a transient time $ \frac{Q}{\omega_0} $ is then given by
\begin{equation}\label{eq:amplitude_projection}
  A(\omega_0) = \frac{Q}{\omega_0^2} \int_{0}^{2\pi/\omega_0}{\rm d}t \, \frac{\omega_0}{\pi}
  \cos(\omega_0 t)
  \int_\Delta^{L+\Delta} {\rm d}x \, \frac{2}{L} \cos \left(\frac{\pi}{2L}(x-\Delta)\right) (-\partial_x \varPhi(x,t)),
\end{equation}
where the integration of $ t $ over one mechanical period gives the Fourier component of the driving force corresponding to this mechanical mode. At this point we assumed the pulse to be centered around $ t=0 $ and to be repeating at intervals of $ \frac{2\pi}{\omega_0} $. 

With the periodic Newtonian potential from appendix \ref{appendix:h_reflected_pulse}
\begin{align}
	\varPhi(x,t) & = \frac{4GP_{\rm p}^{\rm cav}}{c^3} \ln (x) \;\textifsym{l|H|l}_\Sigma (t)
	\\
	\implies -\partial_x \varPhi(x,t) & = -\frac{4GP_{\rm p}^{\rm cav}}{c^3}\frac{1}{x}\; \textifsym{l|H|l}_\Sigma(t),
\end{align}
where $ P_{\rm p}^{\rm cav} $ is the pulse power and $ \textifsym{l|H|l}_\Sigma(t) =\sum_n \left[\Theta\left(t-\frac{n2\pi}{\omega_0}-\frac{\tau_{\rm p}}{2}\right) - \Theta\left(t-\frac{n2\pi}{\omega_0}+\frac{\tau_{\rm p}}{2}\right)\right]$ is a sum of rectangular pulses of duration $ \tau_{\rm p} $.
The integral over the oscillation period in eq.\eqref{eq:amplitude_projection} returns
\begin{equation}\label{eq:fouriercomponent}
  \int_{0}^{2\pi/\omega_0}{\rm d}t \, \frac{\omega_0}{\pi}
  \cos (\omega_0 t)\textifsym{l|H|l}_\Sigma(t) = \frac{2}{\pi}\sin \frac{\omega_0 \tau_{\rm p}}{2} \approx
	\begin{cases}
	\frac{\omega_0 \tau_{\rm p}}{\pi} & \text{ for } \tau_{\rm p}{\omega_0}\ll 2
	\\
	 \frac{2}{\pi} & \text{ for } \tau_{\rm p} \approx \frac{\pi}{\omega_0}
	\end{cases}.
\end{equation}
With this, a resonant maximum amplitude of
\begin{align}\label{eq:Aw02}
	A(\omega_0) = & \frac{4GP_{\rm p}^{\rm cav} Q}{\omega_0^2 c^3  } \frac{2}{\pi}\sin\left(\frac{\omega_0\tau_{\rm p}}{2}\right)\int_\Delta^{L+\Delta} {\rm d}x \, \frac{2}{L} \cos \left(\frac{\pi}{2L}(x-\Delta)\right) \frac{1}{x} 
	\\
	\approx & \frac{32 G P_{\rm cav}^{\rm avg} Q}{\pi \omega_0 c^3} \begin{cases}
	\begin{cases}
	\frac{1}{\omega_0 \Delta}  & \text{ for } \tau_{\rm p}\omega_{0}\ll 2
	\\
	\frac{ 2}{\pi \omega_0 \Delta}  & \text{ for } \tau_{\rm p}=\frac{\pi}{\omega_0}
	\end{cases}
	& \text{ and } L\ll \Delta
	\\\\
	\begin{cases}
          \frac{ 1}{c_{\rm s}} \ln \frac{L}{\Delta} & \text{ for }\tau_{\rm p}\omega_{0}\ll 2
	\\
	\frac{ 2}{\pi c_{\rm s}} \ln \frac{L}{\Delta}  & \text{ for } \tau_{\rm p}=\frac{\pi}{\omega_0}
	\end{cases}
	& \text{ and } L\gg \Delta
	\end{cases}\quad,
\end{align}
where $ P_{\rm cav }^{\rm avg}\equiv \frac{\text{energy in the cavity}}{\text{oscillation period}} = \frac{P_{\rm p}^{\rm cav} \tau_{\rm p}}{\frac{2\pi}{\omega_0}} $ is the power in the cavity averaged over one mechanical period, is reached in the steady state of prolonged driving.   The logarithmic divergence of equation (\ref{eq:Aw02}) for $L\gg \Delta$ is an artifact of idealizations of our model and will not be relevant in practice
\footnote{
  Even though it might seem favorable to increase $ L $ because of the logarithmic scaling in eq. \eqref{eq:Aw02}, the inhomogeneous driving force leads to excitations of multiple mechanical modes which, for a non perfectly rigid support, can couple. Also, the distance between the detector and the beam has to be much smaller than the length of the source cavity (or radius of the ring resonator) for the contributions of the recoil of the mirrors (or deflecting magnetic fields) to be negligible. These two effects might lead to the break-down of the logarithmic scaling before it makes a difference.
  }.

Assuming for orientation numerical values of 
aluminum, 
$c_{\rm s} = c_{\rm s}^{\rm Al}=6420$\,m/s, $ 
\omega_0 = 2 \pi \cdot 10^9 $ Hz, $\Delta = w_{\rm B}$ \footnote{  As the rod was chosen to be fixed by the support at its tip at the far side of the beam, the lowest eigenmode has wavelength $ \lambda = \frac{2\pi c_{\rm s}}{\omega_0} = 4 L $. The minimum distance from the beam is $\Delta \approx w_{\rm B} $, where for the purposes of this estimation, we saturate this lower bound on $ \Delta $.}, and $ Q=10^6 $ for the rod, the laser
cavity introduced in \ref{sec:pulse_cavity} ($ P_{\rm avg} = 20$ GW, $
w_{\rm B} \approx 100 \; \mu$m) would result in an amplitude of
$A\approx  10^{-34}\,$m at the freely oscillating tip. 

At resonance, the noise spectral density for a resonant-bar type detector is given by
\begin{equation}\label{eq:rodthermal}
	S_A^{\rm th} = \frac{4 k_B TQ}{\omega_0^3 M_{\rm eff}} \implies A_{\rm th} = \sqrt{\frac{S_A^{\rm th}}{\tau_{\rm int}}}
\end{equation}
according to \cite[p.440]{maggiore2008gravitational}, where $ M_{\rm eff}= \int \varrho_m A_R (w_0(x))^2 {\rm d}x $ is the effective mass of the mode with the rod cross-section $ A_R $, and $ A_{\rm th} $ is the amplitude resulting from the thermal noise after integration time $ \tau_{\rm int} $. At $ \omega_0 = 2 \pi \cdot 1 $\,GHz the thermal sensitivity limit for temperatures below $ T = 48 $\,mK  is already below the standard quantum limit (SQL) on noise spectral density for a resonant mass detector \cite{clerk2010introduction},
\begin{equation}\label{eq:SQLnoise}
  	S_A^{\rm SQL} = \frac{4 \hbar Q}{M_{\rm eff} \omega_0^2}.
\end{equation}
At frequencies below the megahertz range, the thermal noise 
is the limiting factor.
For $ Q = 10^6 $,  $ M_{\rm eff} = \frac{\pi}{8} \varrho_{\rm Al} L^3  $ (assuming a constant aspect ratio) with the mass density of aluminum $ \varrho_{\rm Al} = 2.7 \,{\rm g/cm^3}$ and a frequency of $ \omega_0 = 2\pi \cdot 1 $\,GHz the sensitivity is 
$\sqrt{S_A^{\rm SQL}} \approx 4 \cdot 10 ^{-17} \,\frac{\rm m}{\sqrt{\rm Hz}}, $
meaning that for 1 year of integration time at best an amplitude of $ 10^{-20} $\,m can be detected.

For the LHC, where the rate of bunches passing by is $ \nu =  31.2 $ MHz,
for the purposes of this estimation, we assume that the same amount of protons is split into  88925 bunches instead of 2808 
such that we reach the frequency $ \omega_0 = 2\pi \cdot 1 $ GHz 
while keeping the same average power. With pulses filling half a period, the peak power is $P_{\rm p}^{\rm cav}=2P_{\rm cav}^{\rm avg}$. 
Using the same $ c_{\rm s}=6420 $ m/s and $ Q=10^6 $ and the values of the LHC ($ P_{\rm cav}^{\rm avg} =  3.8 \cdot 10^{12} $ W, $ w_{\rm B} = 16  $ $\mu$m) one would expect the resonant amplitude to be $ A \approx 9 \cdot 10^{-32} {~\rm m} $, which is at least two orders of magnitude larger than that caused by the oscillating 
laser pulse from Sec.\ref{sec:pulse_cavity}.
For higher quality factors $ Q = 10^8 $ amplitudes of $ A \approx 9 
\cdot 10^{-30} {~\rm m} $ might be possible. At far lower 
frequencies, where the limit $L\gg \Delta$ becomes relevant in
eq.\eqref{eq:Aw02},  a lower speed of sound, for example $ c_{\rm s} = 100 $ m/s, is also beneficial. However, one quickly ends up with
a meter long rod, outside the ``close to the beam'' limit, whilst
still not within range of detection. 

To probe the limit $ L\gg\Delta $, we assume $ \omega_0 = 2\pi \cdot 1 $\,kHz, $ Q=10^6 $, $ c_{\rm s} = 6420 $\,m/s implying an extreme $ L \approx 4 $\,km. 
Then, the expected amplitude from the laser pulses in sec. \ref{sec:pulse_cavity} is $ A \approx 2 \cdot 10^{-25} $\,m, 
for the cavity pumped with a modulated cw laser $ A \approx 4 \cdot 10^{-25} $\,m, 
while we expect an amplitude of $ A\approx 8\cdot 10^{-24} $\,m  for the LHC beam (which would have to be modulated to reach such low frequencies). 
Assuming a temperature of $ T=5 $\,mK the sensitivity is $	\sqrt{S_A^{\rm th}} \approx 10^{-17} \,\frac{\rm m}{\sqrt{\rm Hz}},$ leaving the amplitudes still unmeasurable even for unreasonably long integration and rise times and an unreasonable rod length.

\subsection{  Superfluid  Helium detector}\label{sec:liquid_He}
In \cite{singh_detecting_2017} Singh et al. study the acoustic motion of superfluid helium-4 coupled parametrically to a superconducting microwave cavity as a detection scheme for continuous-wave gravitational signals. With few theoretical adaptations the system can be adapted to the near-field case considered here. The very high Q-factors and sensitive microwave transducer means this is essentially a better version of the deformable rod considered in section \ref{sec:mechanical}.
For the ground mode, the system's description can be reduced to a one dimensional problem and treated as in section \ref{sec:mechanical}, but with two fixed ends instead of one. The spatial displacement amplitude is then given by $  w_0 = \sin \left(\frac{\pi}{L}(x-\Delta)\right) $.

The position noise spectral density of the temporal displacement field $\xi$ is given by eq.~\eqref{eq:rodthermal}, when comparing to the result of Singh et al. \cite{singh_detecting_2017} a factor of 2 has to be added to obtain the single sided density ($ \omega_0>0 $).
With the susceptibility on resonance $ \chi = \frac{Q_{\rm He}}{iM_{\rm eff} \omega_0^2} $, this results in a thermal force noise spectral density (on resonance) of
\begin{equation}
S_{FF}^{\rm th} = \vert\chi\vert^{-2} S_{\xi\xi}^{\rm th} = 4k_B T M_{\rm eff} \frac{\omega_0}{Q_{\rm He}}.
\end{equation}
Which implies a lower bound to the detectable force over an integration time $ \tau_{\rm int} $, with $ 2 \sigma $ uncertainty, of
\begin{equation}
\bar F_{\rm min} \approx 2 \sqrt{\frac{S_{FF}^{\rm th}}{\tau_{\rm int}}} = \sqrt{\frac{16 k_B T M_{\rm eff} \omega_0}{\tau_{\rm int}Q_{\rm He}}}.
\end{equation}

The Fourier component of the force corresponding to the considered 
lowest-frequency mode is given by
\begin{equation}\label{eq:Feff}
\bar{F}_{\rm eff} = \vert \chi \vert^{-1} A(\omega_0) =  \frac{16G M_{\rm eff}  P_{\rm p}^{\rm cav}}{\pi L c^3} \sin \left(\frac{\omega_0 \tau_{\rm p}}{2}\right) \int_\Delta^{\Delta+L} \frac{\sin \Big(\frac{\pi (x-\Delta)}{L}\Big)}{x} {\rm d}x \,.
\end{equation}
Note the similarity of the amplitude $A(\omega_0)$ to the case of the mechanical rod detector in equation (\ref{eq:Aw02})
\footnote{
    In contrast to the spatial integral in (\ref{eq:Aw02}), the one in (\ref{eq:Feff}) converges to $ \approx 1.85 $ for $ L\gg \Delta$. This is because of the different in boundary conditions, in particular, the logarithmic dependence stems from the overlap of the mode function with the steep end of the $ \frac{1}{x} $ driving force, whereas the modes of the helium have to vanish at the end of the container. However, the missing logarithmic dependence is basically irrelevant on realistic length scales. Assuming once again a constant aspect ratio, i.e. $ M_{\rm eff} \sim L^3 $, we find a scaling of $ \bar{F}_{\rm min}  \sim L $ and $\bar{F}_{\rm eff}  \sim L^2$, implying that the force should be detectable if $L$ is large enough. However, limitations apply as is discussed in section \ref{sec:mechanical}.
}. Here, both signal and noise are given as a force, for better comparability to \cite{singh_detecting_2017}.

To get a feeling for the orders of magnitude, we
start off with the numbers from the actual experimental setup from { Singh et al.} \cite{singh_detecting_2017}. We set  
$ \tau_{\rm int} = 250 {\rm \, d}, Q_{\rm He} = 6 \cdot 10^{10}, L = 4 \;
{\rm cm}, r=1.8 \;{\rm cm} \text{ (radius)}, c_{\rm s} = 220 {\rm \; m/s}$,
and $ \varrho_{\rm He} = 145 \;{\rm kg/m}^3 $. This  implies $\omega_0 =
2 \pi \cdot 2.8 \;{\rm kHz}$, $M_{\rm eff} = 3 \; {\rm g},\; T = 5 \;
{\rm mK} $. 
This results in a minimum detectable force $	\bar{F}_{\rm min} \approx 4\cdot 10^{-21} {\;\rm N}$.
Choosing the LHC as a source, we assume $ \Delta =w_{\rm B} $ and set the average power to $ P_{\rm p}^{\rm cav} = P^{\rm LHC} = 3.8 \cdot 10^{12} \;{\rm W} $, $ \tau_{\rm p} = \frac{1}{2}\frac{2 \pi}{\omega_0} $, resulting in
$ \bar{F}_{\rm eff} \approx 6.6 \cdot   10^{-24} {\; \rm N}  $.
Going further from the beamline (by less than $ L $) to account for shielding and the Helium container only decreases the effective force slightly (for $ \Delta = 16\;{\rm \mu m} \rightarrow \Delta = 3 \; {\rm cm} $, $ F_{\rm eff} $ decreases by a factor of 4) as there is limited contribution from the liquid Helium at the ends of the container to the ground mode.

Hence, 
at full amplitude and one week of integration time, the 4\,cm prototype detector is lacking about 3.5 of magnitude in sensitivity. 
Under otherwise identical assumptions,
the proposed first generation (0.5 m) detector 
will be about 2.5
orders of magnitude from being sensitive enough to detect the gravitational signal from the LHC.

\subsection{High-Q milligram-scale monolithic pendulum}
In a recent publication, Matsumoto et al.\cite{matsumoto2019demonstration} described the manufacturing of a pendulum and presented its properties. They found it to have a very high quality factor for a small scale system and even higher when combined with an optical spring.
Different from the extended oscillators considered in the earlier
subsections, the pendulum does not rely on the projection of the
gravitational acceleration on an elastic mode but rather on
the gravitational
force on the pendulum mass relative to the support.
A mechanical oscillator has to be of small scale to be close enough to the
source for the gravitational acceleration to be significant, while
 the gravitational effects on the  pivot point need to remain
negligible. 

For the $l = 1 \;{\rm cm}, m = 7 \;{\rm mg} $ pendulum a mechanical Q-factor of $ Q_{\rm m} = 10^5 $ was measured in \cite{matsumoto2019demonstration} at $ \omega_{\rm m} = 2 \pi \cdot 4.4 \;{\rm Hz} $. Introducing an optical spring to shift the frequency, the effective Q-factor is expected to scale as
\begin{equation}
  \label{eq:Qscaling}
 Q_{\rm eff} \approx Q_{\rm m} \left(\frac{\omega_0}{\omega_{\rm m}}\right)^2
\end{equation}
 for the damping model considered relevant for the pendulum (the effective frequency of the coupled system was renamed from $ \omega_{0} $ ($ Q_0 $) in the original work \cite{matsumoto2019demonstration} to $ \omega_{\rm m} $ ($ Q_{\rm m} $) for consistency).
An additional feedback cooling is necessary to stabilize and cool the system to a temperature $T_{\rm fb}$, compensating the effect of heating through the optical spring. This reduces the $Q$-factor to $Q_{\rm fb}$, which has the benefit of allowing shorter driving times. 
At $ \omega_0 = 2 \pi \cdot 280$~Hz the authors of \cite{matsumoto2019demonstration} demonstrated a sensitivity of $ 3 \cdot 10^{-14} ~\frac{\rm m}{\sqrt{\rm Hz}}$ with a $ Q $-factor of $ Q_{\rm fb} =250$, with thermal motion the main source of noise. According to eq. \eqref{eq:rodthermal} this corresponds to a temperature of a few Millikelvin. 

In an update to this Cata\~no-Lopez et al.\cite{catano-lopez_high-2020} described an improved version of this pendulum, with a measured mechanical Q-factor of $ Q_{\rm m} = 2 \cdot 10^6 $ at a frequency of $ \omega_{\rm m} = 2\pi \cdot 2.2 $~Hz, which with the optical spring is  tunable in the frequency range of $ 400 \;{\rm Hz} < \frac{\omega_0}{2 \pi} < 1800\;{\rm Hz} $.

For a pulsed-beam source, the gravitational acceleration in radial direction for the duration of a pulse is given by
\begin{equation}
a_{\rm grav}^{\rm p} = - \frac{4 GP_{\rm p}^{\rm cav}}{c^3}\frac{1}{\rho},
\end{equation}
where $G$ is the Newton gravitational constant, $ P_{\rm p}^{\rm cav} $ is the pulse power, and $ \rho $ is the distance from the beam.
For this setup $ \rho $ is limited by the radius of the pendulum mass (1.5 mm) and the beam width ($ \ll .5 $ mm), so $ \rho = 2 $ mm is a reasonable estimate which might be substantially increased, however, if a cryostat is needed.

The displacement resulting from prolonged ($ \tau \sim \frac{2\pi Q_{\rm fb}}{\omega_0} $) driving on resonance is given by
\begin{equation}\label{xgrav}
x_{\rm grav} = \frac{\bar{a}_{\rm grav} }{\omega_0^2} Q_{\rm fb}= \frac{8GP_{\rm p}^{\rm cav} \sin \left(\frac{\omega_0 \tau_{\rm p}}{2}\right) Q_{\rm fb} }{\pi c^3 \omega_{0}^2}\frac{1}{\rho},
\end{equation}
where $\bar{a}_{\rm grav}$ is the Fourier component of $ a_{\rm grav}(t) $ 
on resonance, 
and $  \sin \left(\frac{\omega_0 \tau_{\rm p}}{2}\right) $
results from the overlap of the rectangular pulses with the
sinusoidal oscillation calculated in eq.\eqref{eq:fouriercomponent}. We now consider how the pendulum would react to the different gravitational sources discussed above.

\subsubsection{Cavity pumped with cw laser}
For the pendulum from \cite{matsumoto2019demonstration} at $\omega_0 = 2 \pi \cdot 280$ Hz and the cw laser cavity from \ref{sec:cw_cavity} with a power in the cavity of $P_{\rm p}^{\rm cav} = 200 $ GW for half the oscillation period the expected amplitude resulting from the gravitational signal is $ x_{\rm grav}\approx 3.1 \cdot 10^{-26} $ m.
With the on-resonance SQL and thermal sensitivities \cite{clerk2010introduction} $ \sqrt{S_{\rm SQL}} = 2 x_{\rm zpf }\sqrt{\frac{Q_{\rm fb}}{\omega_0}} =\sqrt{ \frac{4 \hbar Q_{\rm fb}}{m \omega_0^2}} \approx 7 \cdot 10^{-17}\;\frac{\rm m}{\sqrt{\rm Hz}}$ and $ \sqrt{S_{\rm th}} = 2 \cdot 10^{-14}\,{\rm m}/\sqrt{\rm Hz}$ (starting from room temperature, with only feedback cooling) this signal amplitude is not measurable.

The SQL refers to amplitude-and-phase measurements of that position. 
In principle, due to the precisely known frequency, quantum non-demolition
measurements allow continuous monitoring of the oscillation \cite{Caves1980}.
With a ``single-transducer, back-action evading measurement'', one can estimate a quadrature of the oscillator with an uncertainty that scales $\propto (\beta\omega_0\tau_{\rm m})^{-1/2}$, where $\tau_{\rm m}$ is the relevant measurement time or inverse filter width, and $\beta$ a numerical factor that can reach a value of order one (see eq.~3.21a,b in \cite{Caves1980} and eqs.(32,33) in \cite{braginsky_quantum_1980}).  
After upconverting the kHz signal to the GHz regime one can use modern microwave amplifiers with essentially no added noise \cite{PhysRevX.6.041024,PhysRevLett.118.103601,zhong_squeezing_2013,toth_dissipative_2017}.  Upconversion to the microwave frequency range was already discussed in the 1980s \cite{braginsky_quantum_1980} and can be achieved by having the sensor modulate the resonance frequency of a microwave cavity.
Additional sensitivity can be gained with a large number $N$ of sensors arranged along the laser beam or particle beam. Classical averaging their signal leads to a noise reduction of $1/\sqrt{N}$ in the standard deviation. When several sensors all couple to the same microwave cavity, one might even hope to achieve ``coherent averaging'', in which case the noise reduction scales as $1/N$ \cite{fraisse_coherent_2015,braun_coherently_2014}. 

With $N=1$ and a signal of $ 280 $\,Hz, 
the sensitivity of the pendulum resulting from the standard quantum noise limit $ \sqrt{S_{\rm SQL}} \approx 7 \cdot 10^{-17} \,\frac{\rm m}{\sqrt{\rm Hz}} $ is 
3 orders of magnitude lower than that given by the thermal noise. For 1 week of measurement, the thermal noise still exceeds the signal generated by the modulated cw laser (respectively train of laser pulses) by 8 (almost 9) orders of magnitude.

\subsubsection{LHC beam}\label{sec:LHC}
The minimum frequency of one bunch of ultra-relativistic protons going
around the ring of the LHC is in the kHz range 
(see Sec.\ref{Sec.LHC1}). Lower frequencies could be achieved by modulating the beam position with low frequency.  
The LHC as a source is expected 
to create almost $ 20 $ times larger amplitude than the considered 
cw-pumped cavity, due to the higher pulse power $ P_{\rm p}^{\rm cav} = P^{\rm LHC} $ where an ``on-off'' modulation of the LHC beam, similar to the cw cavity pumping scheme was assumed.
After one week of measurement time one would be about 
7 orders of magnitude off from measuring the signal with a single detector, 5 orders of magnitude starting at a temperature of $ T = 5 $\, mK.  
Substantially more development will be needed 
to bridge this gap. Ideas in this direction are developed in the next section.

\subsection{Optimizing the S/N} \label{sec.optimization}
In this section we ask, what parameter values would be needed to
achieve a signal-to-noise ratio comparable to 1 for a torsion balance
or pendulum.  We model both simply as damped harmonic oscillators, but
keep in mind that their mechanial parameters and temperature can be
substantially modified by using an optical spring and/or feedback
cooling, and then compare to the existing setups described in
\cite{westphal_measurement_2020,matsumoto2019demonstration,catano-lopez_high-2020}. We
therefore continue to use $T_{\rm fb}$ for the final temperature,
$Q_{\rm fb}$ for the final quality factor, and $\omega_0$ as final
resonance frequency $\times 2\pi$, regardless of how they might be
achieved.\\ 

\subsubsection{Optimization of a mechanical oscillator as dectector and comparison to \cite{westphal_measurement_2020}}\label{sec.opt}
According to eq.(5.60) in \cite{clerk2010introduction}  the total position-noise power at frequency $\omega$ of a harmonic oscillator with (undamped) resonance frequency $\Omega$ measured with a 
transducer 
and amplifier that add back-action noise (referred back to the input) can be written as
\begin{equation}
  \label{eq:Sxxtot}
  \bar{S}_\text{xx,tot}(T,\omega,\Omega,Q,m)=\frac{\gamma_0}{\gamma_0+\gamma}\bar{S}_\text{xx,eq}(T,\omega,\Omega,Q,m)+\bar{S}_\text{xx,add}(T,\omega,\Omega,Q,m)\,
\end{equation}
where $\gamma_0$ is the intrinsic oscillator damping without coupling to the transducer 
and $\gamma\equiv\gamma(\omega)$ the damping with the 
coupling. The equilibrium noise (comprising both quantum noise and thermal noise at temperature $T$) reads
\begin{eqnarray}
  \label{eq:Sxxeq}
  \bar{S}_\text{xx,eq}(T,\omega,\Omega,Q,m)&=&\hbar \, {\coth}\left(\frac{\hbar \omega}{2k_B T}\right) \text{Im} \chi_\text{xx}(\omega,\Omega,Q,m)\\
 \text{Im} \chi_\text{xx}(\omega,\Omega,Q,m)&=&\frac{Q \omega \Omega}{m(\omega^2\Omega^2+Q^2 (\omega^2-\Omega^2)^2)}\,,
\end{eqnarray}
with the quality factor $Q\equiv \Omega/(\gamma_0+\gamma)$. To calculate $\bar{S}_\text{xx,add}$, one needs to know the force noise power of the detector and amplifier, but $\bar{S}_\text{xx,add}$ is lower bounded by $\bar{S}_\text{xx,addMin}=\hbar |\text{Im} \chi_\text{xx} | $. With this lowest possible value and the replacements $\Omega\to \omega_0$, $Q\to Q_\text{fb}$, $ T\to T_{\rm fb} $, one obtains for the total noise power to lowest order in $\gamma/\gamma_0$ (which slightly overestimates the contribution from $\bar{S}_\text{xx,eq}(T,\omega,\omega_0,Q,m)$)
\begin{equation}
  \label{eq:Sxxfin}
  \bar{S}_\text{xx,tot}=\hbar\left(1+\coth\left(\frac{\hbar \omega}{2k_B T_{\rm fb}}\right)\right) \text{Im} \chi_\text{xx}(\omega,\omega_{0},Q_\text{fb},m)\,.
\end{equation}
The maximum amplitude $x_\text{grav}$ of the harmonic oscillator is
given by eq.\eqref{xgrav} with $\sin(\tau_{\rm p}\omega_0/2)=1$, but
is reached only asymptotically as function of time, namely as
$x_\text{grav}(t)=x_\text{grav}(1-\exp(-\omega_0 t/(2Q_\text{fb})))$.
We assume that the total time $\tau_\text{tot}=1$ week for the
experiment is split as $\tau_\text{tot}=\tau_{\rm r}+\tau_{\rm m}$
into a time $\tau_{\rm r}$ needed for the amplitude of the harmonic oscillator to rise to a certain level, and a measurement time $\tau_{\rm m}$ used for reducing the noise. The total signal-to-noise ratio on resonance is then given by
\begin{eqnarray}
  \label{eq:ston}
  S/N&=& x_\text{grav}\left(1-\exp(-\omega_0 (\tau_\text{tot}-\tau_{\rm m})/(2Q_\text{fb}))\right)\frac{\sqrt{\tau_{\rm m}}}{\sqrt{\bar{S}_\text{xx,tot}}}\nonumber\\
  &\simeq& 0.01 \, \frac{(1-e^{((\tau_{\rm m}-\tau_\text{tot})\frac{\omega_{0}}{2Q_{\rm fb}}})\sqrt{Q_{\rm fb} m \,\tau_{\rm m}}}{\omega_{0}\sqrt{1+\coth\frac{4\cdot 10^{-12} \omega_0}{T_{\rm fb}}}}\,,\label{eq:ston2}
\end{eqnarray}
where a distance $\rho=200\, \mu$m of the center of the detector mass 
from the beam axis was assumed. All quantities are in SI units. From this equation it is evident that the mass $m$ should be as large as possible.  At the same time, $m$ cannot be made arbitrarily large, as otherwise the distance from the beam axis would have to be increased as well, which would lead to a decay of the signal $\propto 1/\rho$ for $\rho\gg \rho_\text{min}$, where $\rho_\text{min}$ is the minimum distance from the beam axis (which might contain a shielding of the particle beam in the case of the LHC, and which we assume to be $\rho_\text{min}=100\,\mu$m for the LHC but might have to be substantially increased when using a cryostat).
In principle, for a spherical detector mass, a scaling $\propto
m^{1/6}$ would still result, but it turns out that unrealistically
large masses (larger than 1\,kg) would be needed before this scaling
gives an advantage over an alternative design with a cylindrical
detector mass that allows to maintain $\rho=200 \,\mu$m.  If we allow
that cylinder to become as long as $L_\text{cyl}=0.5$\,m and determine
the maximum mass as $m=0.9 \pi \varrho_\text{Si}
(\rho-\rho_\text{min})^2\,L_\text{cyl}$ (where 0.9 is a ``fudge
factor'' that avoids that the detector mass touches the shielding), we find $m=33$\,mg. \\

With that value inserted in eq.\eqref{eq:ston2}, one can optimize
$S/N$ with respect to the parameters $\tau_{\rm m},\omega_{0},Q_{\rm
  fb}$ and $T_{\rm fb}$.
With $\tau_{\rm tot}$ kept equal to 1 week, 
in the range $1$ rad/s $\le \omega_0\le 10^4$ rad/s, $1\le Q_{\rm fb}\le 10^8$, 1\,nK $\le T_{\rm fb}$
 a maximum value $S/N\simeq 0.6$ 
  is found for $\tau_{\rm m}=3\cdot 10^5$\,s, $\omega_0=2\pi \cdot 0.16$ Hz, $ Q_{\rm fb} = 1.2 \cdot 10^5 $, and minimal $T_{\rm fb}$. 
 The optimal value for $\omega_{0}$ is at the lower end of the parameter range, but reasonably close to the one for the existing torsion balance in \cite{westphal_measurement_2020} ($\omega_0=2\pi\times 3.59$\,mHz), where, however, the mechanical quality factor was $Q=4.9$ and a mass of 92.1\,mg was used. It remains to be seen if the parameters that result from the optimization can be reached.  Problematic appears mostly whether the temperature of the cooled mode of about 1\,nK can be reached, especially at low frequencies. 

 \subsubsection{Assumption of $Q$-scaling and comparison to \cite{matsumoto2019demonstration}}
 The structural damping model used in
 \cite{matsumoto2019demonstration} implies a quadratic scaling of the
 Q-factor with the resonance frequency (see
 eq. \eqref{eq:Qscaling}).  Including this scaling
 behavior and allowing the modification of the resonance frequency by means of an optical
 spring, leads to frequencies in the
 100\,Hz to 1\,kHz range being preferred by the optimization. 
This ultra-high $ Q $-factor  is, however, not reachable in practice
as the optical spring introduces heating, and so the mechanical oscillator has to be cooled  to stabilize the system. In existing systems, feedback cooling \cite{PhysRevD.65.042001,matsumoto2019demonstration}, or a second optical spring tuned to the infrared \cite{PhysRevD.78.062003}  have been employed as cooling mechanisms. We assume an effective final temperature reached by feedback cooling of
\begin{equation}
	T_{\rm fb} = 4 T_{\rm bath}\frac{Q_{\rm fb}}{Q_{\rm eff}}=  4 T_{\rm bath}\frac{Q_{\rm fb}}{Q_{\rm m} \frac{\omega_{0}^2}{\omega_{\rm m}^2}},
\end{equation}
as is expected in \cite{matsumoto2019demonstration}. 
An initial temperature, before feedback cooling, of $T_{\rm bath}= 5$\,mK is assumed and the parameter ranges are limited to $ 1$\,rad/s $<\omega_{\rm m}<10000$ rad/s, $ 1$ rad/s $<\omega_0 <10000$ rad/s, $1< Q_{\rm m}<10^7 $, and $1< Q_{\rm fb}< 10^{10}$. We find $S/N\approx0.077$ for the optimal parameters of $\omega_{\rm m} = 2\pi\cdot 0.16$\,Hz, $\omega_{0} = 2\pi\cdot 600$\,Hz, $Q_m = 10^7$, and $Q_{\rm fb} = 1.6 \cdot 10^8$.
Compared to the generic optimization as seen above this seems underwhelming but if the scaling of $Q$ and temperature can be attained, the final temperature of $ T_{\rm fb} \approx 23 $\,nK would be more feasible than before.						
 
\subsubsection{Further possible improvements}
A signal-to-noise ratio of $0.6$ is still not good enough, but the
planned upgrade of the LHC to the high-luminosity LHC
\cite{krasny_high-luminosity_2020} should increase $S/N$ by a factor
10. Another factor 2.9 is expected to be gained by switching to
tungsten (with mass density $\varrho_{\rm W}=19,250$\,kg/m$^3$) as
detector-mass material, all other optimized parameters remaining equal. Both factors combined lead to a $S/N\simeq 16$. \\

The maximum of $S/N$ found in the optimization is rather flat, especially with respect to the feedback cooling quality factor, such that there is a wide range of values with similar signal-to-noise ratios that allow one to take into account other engineering constraints not considered here and without such extreme effective temperatures. 
Hence, with the high-luminosity LHC and an optimized detector there is
realistic hope that GR could be tested for the first time in this
ultra-relativistic regime with a controlled terrestrial source and
adapted optimized detector.  Also without the upgrade of the LHC, further
improvements from using a multitude of detectors (and possibly
coherent averaging by coupling them all to the same read-out cavity \cite{fraisse_coherent_2015,braun_coherently_2014}) or longer integration times can be envisaged that would bring $S/N$ to order one. 
\begin{table}
	\begin{tabular}{|c|c|c|c|}
		\hline
		& rod & liquid helium & pendulum\\
		\hline
		$ \omega_0 $ &  $ 2 \pi \cdot 10^3 $ Hz \vline\, $2 \pi \cdot 10^9 $ Hz & $2\pi \cdot 2.8 \cdot 10^{3} $ Hz & $2 \pi \cdot 280 $ Hz 
		\\
		\hline
		sensitivity  
		& $ 1 \cdot 10^{-17} $\,$\frac{\rm m}{\sqrt{\rm Hz}}$
		\vline\, $4 \cdot 10^{-17}$\,$\frac{\rm m}{\sqrt{\rm Hz}}$
		&$ 2 \cdot 10^{-17}  $\,$\frac{\rm N}{\sqrt{\rm Hz}}$ \vline\, $1 \cdot 10^{-12} $\,$\frac{\rm m}{\sqrt{\rm Hz}}$
		&  $ 2 \cdot 10^{-14}$\,$\frac{\rm m}{\sqrt{\rm Hz}}$ 
		\\
		\hline
		limiting factor & thermal noise \vline\,\;\qquad SQL\qquad\qquad  & thermal noise & thermal noise \\
		\hline
		expected amplitude\\\hline\hline
		laser pulses & $ 2 \cdot 10^{-25} $ m \vline\,$ 2\cdot 10^{-34} $ m &$ 2 \cdot 10^{-25} $ N \vline\,  $ 1 \cdot 10^{-20} $ m & $ 3 \cdot 10^{-26} $ m
		\\\hline
		cw cavity & $ 4 \cdot 10^{-25} $ m \vline\,$  4\cdot 10^{-34} $ m & $ 3 \cdot 10^{-25} $ N \vline\, $ 2 \cdot 10^{-20} $ m & $ 5 \cdot 10^{-26} $ m
		\\\hline
		LHC beam ${}^\ast$ & $ 8 \cdot10^{-24} $ m \vline\,$ 9 \cdot  10^{-32} $ m & $ 7 \cdot 10^{-24} $ N \vline\, $ 4 \cdot 10^{-19}$ m & $  1 \cdot 10^{-24} $ m\\
		\hline

	\end{tabular}
	\caption{Comparison of the estimated sensitivity of the listed detectors with the expected amplitude of the sources considered 
          on resonance and after the full build-up-time of the detector's oscillation. For the cases in which the main limiting factor is thermal noise, a temperature of 5\,mK was assumed. 
          Other parameters see text. \\  
          ${}^\ast$ assuming the LHC beam can be modulated to produce a signal with appropriate frequency while maintaining the same average power.}
\end{table}

\section{Discussion}\label{Sec.discus}

\subsection{Perspectives for measuring the gravitation of light or
  particle beams}
We have theoretically investigated the fundamental limitations to measure the oscillating gravitational fields of lab-scale ultrarelativistic sources for three concrete examples: for laser beams, we have considered femtosecond-pulse lasers fed into a high finesse cavity, where they oscillate to and fro, and similarly, cw lasers used to pump a cavity periodically.  For particle beams, we considered the LHC with its beam of proton bunches flying along the accelerator ring. All sources considered lead to oscillating curvature of space-time and acceleration of test particles with precisely controlled frequency up to the GHz range. In addition, we have given details on how modulations of these signals with much lower frequency, down to the kHz regime, can be achieved for all three example sources. 
In the latter regime, the LHC is the most promising ultrarelativistic source of gravity with a gravitational field strength 20 times stronger than the laser sources considered here.

We investigated three near-field detectors:
A deformable rod offers force accumulation along its length thanks to its Young modulus. However, the spatial decrease of the studied gravitational effects limits the effects of force accumulation, resulting in immeasurably small amplitudes of the order of $ 8\cdot 10^{-24} $\,m even in the case of the LHC as a source. 
In the liquid helium chamber from { Singh et al.} \cite{singh_detecting_2017}, very high quality factors and low noise allow for sound wave buildup within the chamber. With the  present experimental parameters \cite{singh_detecting_2017}, the gravitational force for the LHC is 3.5 orders of magnitude below the detectability limit of this detector with an averaging time of one week.

A pendulum from \cite{catano-lopez_high-2020} and
\cite{matsumoto2019demonstration} or a recently demonstrated torsion balance
\cite{westphal_measurement_2020} turned out 
to be the most promising detectors.  
In the present form of the monolithic pendulum
\cite{matsumoto2019demonstration}, the fully built-up signal from the LHC is 5 orders 
of magnitude away from the sensitivity achievable within 1 week of
averaging time (assuming a starting temperature of $ T=5 $\,mK and a final temperature of $ T_{\rm fb} = 12 $\,nK after a shift of the resonance frequency via an optical spring and feedback cooling) with the benefit of a relatively small signal rise time.

Optimization of the signal-to-noise ratio of a mechanical oscillator
as detector over its frequency, measurement time, quality factor and
temperature in the parameter range provided in Sec.\ref{sec.opt},
leads to an expectation of a $S/N$ of about 0.6 with the LHC as source within one week of
signal rise time and averaging. By using a denser material such as
tungsten for the detector mass and profiting from the planned
high-luminosity upgrade of the LHC a S/N ratio $\simeq 16$ appears
possible with one week of measurement time for a single detector. 

Our considerations concerned fundamental limitations so far, so that a S/N ratio larger than 1 should be considered a necessary condition, but would still make for a very difficult experiment with additional noise  and engineering issues to be overcome (see e.g. \cite{schmole_micromechanical_2016}).

Important additional noise sources that have not been considered in our work include, for example, seismic and thermal noise that may be reduced by moving to a higher frequency regime. 
Therefore, while very high source frequencies (GHz) turned out to be detrimental for the considered detectors, 
it may still be interesting to investigate an intermediate frequency range above the kHz regime. 
In their current design, the superfluid helium detector \cite{singh_detecting_2017} and the pendulum detector \cite{catano-lopez_high-2020} need a source oscillating with a frequency of the order of kHz and 400 Hz to kHz, respectively. The pendulum's operation at higher frequencies might be possible and relatively easy to achieve, given that the relevant noise terms in the kHz range stem from suspension eigenmodes, which are changeable by design. Also in \cite{hartwig2020mechanical} parametric cooling into the ground state for pendulum-style gravitational sensors was demonstrated, reducing problems from thermal noise and seismic noise in an even larger frequency range. However, reaching the required low-temperatures in the nK regime in combination with the high quality factors will remain a huge challenge, even if the $Q$-scaling \eqref{eq:Qscaling} and feedback cooling assumed in \cite{matsumoto2019demonstration} is achieved.

\subsection{Perspectives for quantum gravity experiments}\label{sec.qmgr}
The realization that the gravitational effect of light or high-energy
particle beams might become measurable in the near future opens new
experimental routes to quantum gravity, in the sense that it might
become possible to study gravity of light or matter in a non-trivial
quantum superposition.  Concerning light, non-classical states of
light, in particular in the 
form of squeezed light, have been studied and experimentally realized for a
long time, and are now 
used for enhanced
gravitational-wave-sensing in LIGO and Virgo \cite{aasi_enhanced_2013,acernese2019increasing}.  While the current records of
squeezing were obtained for smaller intensities than relevant for
the gravitational sources we consider here
\cite{vahlbruch_detection_2016}, squeezing  and entanglement shared by many
modes was already achieved for photon numbers on the order $10^{16}$
by using a coherent state in one of the modes \cite{Keller08}.  This 
is substantially smaller than the $\sim 10^{21}$ photons estimated in
the cavity in the example of the cw laser leading to 100\,GW 
circulating power considered above, but
one might hope that technology progresses to achieve at
least a small
amount of squeezing also for the high-power sources relevant here. 

As for the high-energy particle beams, transverse ``coherent oscillations'' of
two colliding accelerator beams (including the ones at LHC) have
already been studied
\cite{alexahin_landau_1996,buffat_coherent_2014,alexahin_coherent_2001,zorzano_coherent_1999,yokoya_tune_1990}
but these are of classical nature. Non-trivial quantum states of the
beam are those that cannot be described by a positive semidefinite
Glauber-Sudarshan $P$-function, a concept from quantum optics that is
well established for harmonic oscillators and is 
hence applicable to small-amplitude transverse motion of the particle beam in the
focusing regions where there is a linear restoring force. 
A stronger requirement
would be a non-positive-semidefinite Wigner function, which can be
applied to any system with a phase space.  In order to reach such
quantum states, it will be necessary to cool the particle beams.
Efforts to do so are on the way or proposed for other motivations:
cooling enhances the  phase space density and hence the intensity of the
beam in its center. In addition, new phases of matter in the form of
classical crystalline beams attracted both theoretical
and experimental interest  at least since the 1980s \cite{schiffer_could_1985,schatz_crystalline_2001,wei_crystalline_beams_2003,tanabe_longitudinal_2008,schramm_bunched_2001,schramm_cooling_2003}.  Recently it was proposed to
extend this work to create an ``ultracold crystalline beam''
and turning an ion beam into a quantum computer. For this,
the beam should become an ion Coulomb crystal cooled to
such low temperatures that the de Broglie wavelength becomes larger
than the particles' thermal oscillation amplitudes \cite{brown_towards_2020}. Ideally, for our purposes, the
center-of-mass motion of the beam should be cooled to the ground state
of the (approximate) harmonic oscillator that restrains locally, at the detector
position, the transverse motion, before interesting quantum
superpositions can be achieved. 

However, even superpositions in longitudinal direction would create an interesting experimental
situation for which there is currently no theoretical prediction.   Experimental progress in this direction would allow a different kind of search for quantum gravity effects compared to popular current attempts to detect deviations from canonical commutation relations between conjugate observables as predicted by various quantum-gravity candidates (see e.g.~\cite{vasileiou_planck-scale_2015}). 
Different techniques for cooling particle beams are available (see e.g.~\cite{sessler_cooling_1996,steck_cooling_2015} for overviews): 
Stochastic cooling (measurement of 
deviation from the ideal beam-line and fast electronic
counter-measures further down the beamline) was used at CERN for
producing high-intensity anti-proton beams from 1972 till 2017 and is
still used there for anti-proton deceleration, as well as 
at Forschungszentrum J\"ulich (COSY experiment) and GSI
Helmholtzzentrum f\"ur Schwerionenforschung GmbH (Heavy Ion storage
ring ESR); electron cooling (absorption of entropy by a co-propagating
electron beam of much lower energy and entropy), and a modern cousin of it,
``coherent electron cooling'' \cite{litvinenko_coherent_2009}, under
development at Brookhaven 
National Lab for ion energies up to 40\,GeV/u for Au$^{+79}$ ions \cite{litvinenko_plasma-cascade_2018,litvinenko_coherent_2017}; and
laser cooling, with which longitudinal temperatures on the order
of mK have been reached for moderately relativistic ion beams
\cite{hangst_laser_1991,schroder_first_1990}.  Laser cooling is most
efficient for longitudinal cooling, 
but transverse cooling can be achieved to some extent through
sympathetic cooling \cite{miesner_efficient_1996}. Laser cooling is now proposed for an
ultrarelativistic heavy-ion upgrade of the LHC \cite{krasny_high-luminosity_2020}.  
Despite all these techniques, ground states of the transverse center-of-mass
motion have never been reached in any ultra-relativistic particle beam as far as we know,
nor was it perceived as an important goal. We hope that the
perspective of winning the race to the first quantum gravity
experiment will change this. As Grishchuk put it \cite{grishchuk_electromagnetic_2003}: ``The
laboratory experiment is bound to be expensive, but one should remember that a part of the cost is
likely to be reimbursed from the Nobel prize money !{}``.  The successful
development of ion-trap quantum computers, where ground-state cooling
of collective modes of ion crystals has become standard, might lend
credibility to the feasibility of the endeavor.
\\
\\
{\em Acknowledgments:} We thank Daniel Est\`eve for discussion,
  correspondence and references,  and for proposing the idea to look
  at pulses in a cavity; Werner Vogelsang for a discussion on particle accelerator beams, Nobuyuki Matsumoto and Eddy Collin for correspondence.   DR acknowledges funding by the Marie Skłodowska-Curie Action IF program -- Project-Name "Phononic Quantum Sensors for Gravity" (PhoQuS-G) -- Grant-Number 832250.
  We acknowledge support by Open Access Publishing Fund of University of T\"ubingen.

\appendix

\section{Gravitational field of a laser-pulse in a cavity}\label{appendix:h_reflected_pulse}
Following the calculation of the gravitational field of a box shaped laser pulse of length $ L $ emitted at $ z=0 $ and absorbed at $ z=D $ from \cite{ratzel_gravitational_2016,tolman_1931}, we extend the calculation to an oscillation of a short pulse ($ L<D $) between $ 0 $ and $ D $.
For a pulse propagating along the $\pm z $-direction the energy momentum tensor is given by $ T_{00}=T_{zz} = \mp T_{0z}=\mp T_{z0} = u(z,t)\delta(x)\delta(y)A $, where $ u(z,t) $ is the energy density of the electromagnetic field in 3D and $ A $ is the effective transverse area.
This energy momentum tensor violates the continuity equation as the
recoil of the mirrors is neglected. However, ultimately only positions
very close to the beam will be considered 
where these contributions 
vanish \cite{tolman_1931}.
The energy density is given by 
\begin{equation}
u(z,t) = u_{\rm p}  \Theta(z)\Theta(D-z) \sum_{n=0}^\infty \big(\chi^n_+(z,t)+\chi^n_-(z,t)\big),
\end{equation} where
\begin{align}
\chi^n_+(z,t) = &\Big(\Theta \big(ct-2nD-z\big)-\Theta\big(ct-2nD-z-L\big)\Big)
\\
\chi^n_-(z,t) = &\Big(\Theta \big(ct-(2n+1)D+(z-D)\big)-\Theta\big(ct-(2n+1)D+(z-D)-L\big)\Big)
\end{align}
delimit the profile of the pulse injected at $ t=0$ and reflected $ 2n $ times ($ 2n+1 $ times) traveling in positive (negative ) $ z $-direction, and $ u_{\rm p} = \frac{E_{\rm p}}{L A} $ is the pulse energy density.

From the wave equation 
in the Lorenz gauge 
\begin{equation}
\Box h_{\mu\nu} = -\frac{16\pi G}{c^4} T_{\mu\nu}
\end{equation}
the metric perturbation can be calculated using the Green's function
\begin{equation}
h_{\mu\nu} ( \vec r, t) = \frac{4G}{c^4} \int {\rm d}^3 r' \frac{T_{\mu\nu} (\vec r',t-|\vec r-\vec r'|/c)}{|\vec r - \vec r'|}.
\end{equation}

Given the energy-momentum tensor, the metric perturbation can be decomposed into $ h_{\mu\nu} =  h^+_{\mu\nu} + h^-_{\mu\nu}$ and the only non-zero elements of $ h^{\pm}_{\mu\nu} $ are
$ h^\pm_{00} = h^\pm_{zz} = \mp h^\pm_{z0} = \mp h^\pm_{0z} \equiv h^\pm$, with
\begin{equation}
h^\pm (x,y,z,t) =  \frac{4G u_{\rm p} A}{c^4} \int_{0}^{D}{\rm d}z' \frac{\sum_n \chi_\pm^n\big(z',t-\sqrt{\rho^2+(z-z')^2}/c\big)}{\sqrt{\rho^2+(z-z')^2}},
\end{equation}
$ \rho = \sqrt{x^2+y^2} $, and $ u_{\rm p} A = \frac{P}{c} $.

The box function $ \chi_+^n $ imposes the additional boundaries of $ a_+^n< z'<b_+^n $, with
\begin{align}
a_+^n = & z + \frac{(ct-2nD-L-z)^2-\rho^2}{2(ct-2nD-L-z)}
\\
b_+^n = & z + \frac{(ct-2nD-z)^2-\rho^2}{2(ct-2nD-z)}.
\end{align}
Similarly, the box function $ \chi_-^n $ adds the constraints of $ a_-^n < z' < b_-^n $, with
\begin{align}
b_-^n = & z-\frac{\big(ct-(2n+1)D -L +(z-D)\big)^2 - \rho^2}{2\big(ct-(2n+1)D -L +(z-D)\big)}
\\
a_-^n = & z-\frac{\big(ct-(2n+1)D +(z-D)\big)^2 - \rho^2}{2\big(ct-(2n+1)D  +(z-D)\big)}.
\end{align}

Following \cite{ratzel_gravitational_2016} the substitution $ \zeta (z') = z'-z + \sqrt{\rho^2+(z'-z)^2} $ is used to further simplify the integration. The constraints turn into
\begin{align}\label{eq:zetabound1}
\zeta (0) = & r-z
\\
\zeta (D) = & r_D -(z-D)
\\
\zeta (a_+^n) = & ct-2nD-L-z
\\
\zeta (b_+^n) = & ct - 2nD -z
\\
\zeta (b_-^n) = & \frac{\rho^2}{ct-(2n+1)D - L + (z-D)}
\\
\zeta (a_-^n) = & \frac{\rho^2}{ct-(2n+1)D + (z-D)},\label{eq:zetabound2}
\end{align}
where $ r = \sqrt{\rho^2+z^2} $, and $ r_D = \sqrt{\rho^2+(z-D)^2} $.

\begin{figure}
	\centering
	\def\svgwidth{.4\columnwidth}
	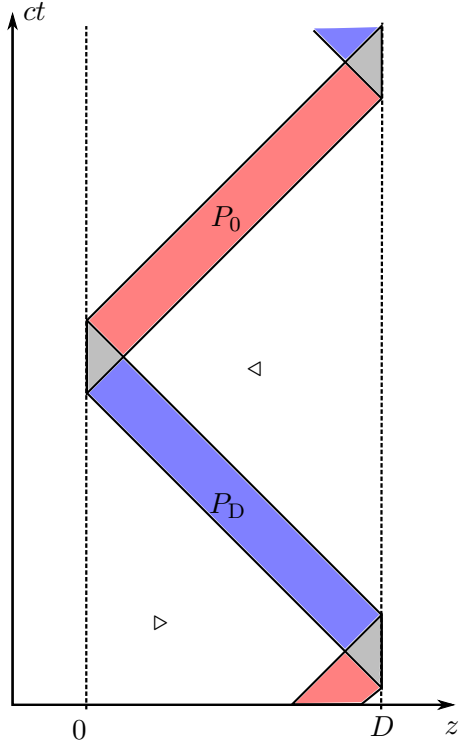
	\caption{Spacetime is split into zones by the light cones of the reflection events of a pulse oscillating between two mirrors.
	}\label{fig:spacetimezones}
\end{figure}

For an observer positioned at $ z\in (L,D-L) $ there are 4 different space-time zones (see fig. \ref{fig:spacetimezones})
\begin{itemize}
	\item[$ P_0^n $:] $ 2nD < ct-r < 2nD+L $ 
	\\
	causally connected to the reflection at $ z=0 , ct = 2nD $
	
	\item[$ \triangleright^n $:] $ 2nD + r +L < ct < (2n+1) D + r_D $ 
	\\
	not causally connected to any reflection events, but causally connected to the pulse traveling from $ z=0, ct = 2nD $ to $ z=D, ct = (2n+1)D $
	
	\item[$ P_D^n $:] $ (2n+1) D < ct- r_D < (2n+1)D+L $
	\\
	causally connected to the reflection at $ z=D, ct = (2n+1)D $
	
	\item[$ \triangleleft^n $:] $ (2n+1)D + r_D + L < ct < (2n+2)D+r $
	\\
	not causally connected to any reflection events, but causally connected to the pulse traveling from $ z=D, ct = (2n+1)D $ to $ z=0, ct = (2n+2)D $.
\end{itemize}  

The metric perturbation is then given by
\begin{equation}\label{eq:hplus}
h^+ = \frac{4GP}{c^5} 
\begin{cases}
\ln \frac{\zeta(b_+^n)}{\zeta(0)}=\ln \frac{ct_{2n}-z}{r-z}   
& \text{ for } (z,t) \in P_0^n,
\\
\ln \frac{\zeta(b_+^n)}{\zeta (a_+^n)} =\ln \frac{ct_{2n} - z}{ ct_{2n}-L-z} 
& \text{ for } (z,t)  \in \triangleright^n,
\\
\ln \frac{\zeta(D)}{\zeta(a_+^n)} = \ln \frac{r_D-(z-D)}{ct_{2n+1} - L - (z-D)}
& \text{ for } (z,t) \in P_D^n,
\\
0
& \text{ for } (z,t) \in \triangleleft^n,
\end{cases}
\end{equation}
caused by the pulses starting from $ z=0 $
and
\begin{equation}\label{eq:hminus}
h^- = \frac{4GP}{c^5} 
\begin{cases}
\ln \frac{\zeta (b_-^{n-1})}{\zeta(0)}=\ln \frac{\rho^2}{(ct_{2n}-L+z)(r-z)}
& \text{ for } (z,t) \in P_0^n,
\\
0 
& \text{ for } (z,t)  \in \triangleright^n,
\\
\ln \frac{\zeta(D)}{\zeta(a_-^n)}=\ln \frac{(r_D-(z-D))(ct_{2n+1} + (z-D))}{\rho^2}
& \text{ for } (z,t) \in P_D^n,
\\
\ln \frac{\zeta(b_-^n)}{\zeta(a_-^n)}=
\ln \frac{ct_{2n+1}  + (z-D)}{ct_{2n+1} -L + (z-D)}
& \text{ for } (z,t) \in \triangleleft^n,
\end{cases}
\end{equation}
caused by the pulses returning from $ z=D $,
where $ t_j := t-jD/c $.

Following \cite{ratzel_gravitational_2016}, the only independent non-vanishing elements of the Riemann curvature tensor are given by
\begin{align}
R_{0z0z} = & -\frac{1}{2}\Big(\frac{1}{c}\partial_t + \partial_z\Big)^2 h^+ -\frac{1}{2}\Big(\frac{1}{c}\partial_t - \partial_z\Big)^2 h^-
\\
R_{0z0i} = & - R_{0zzi} = -\frac{1}{2} \partial_i \Big(\frac{1}{c}\partial_t+\partial_z\Big) h^+ -\frac{1}{2} \partial_i \Big(\frac{1}{c}\partial_t-\partial_z\Big) h^-
\\
R_{0i0j} = & R_{zizj} = -\frac{1}{2} \partial_i\partial_j (h^++h^-)
\\
R_{0izj} = & \frac{1}{2} \partial_i\partial_j ( h^+-h^-),
\end{align}
where $ i,j \in \{x,y\}$.

Given the explicit form of $ h^+ $ and $ h^- $ from eqs.\eqref{eq:hplus} and \eqref{eq:hminus}, the curvature is
\begin{itemize}
	\item[$ P_0^n $:] 
	$ R_{0z0z} = \frac{4GP}{c^5} \frac{z}{r^3} , \quad 
	R_{0z0i}=0, \quad R_{0i0j} = \frac{4GP}{c^5} \frac{z}{\rho^2 r}\Big(\delta_{ij}-\frac{r_i r_j}{\rho^2r^2}(2r^2+\rho^2)\Big), \quad
	R_{0izj} = -\frac{4GP}{c^5}\frac{1}{\rho^2}\left(\delta_{ij}-\frac{2 r_i r_j}{\rho^2}\right) $
	
	\item[$ \triangleright^n $:] $ R_{\mu\nu\rho\sigma} = 0 \quad\forall \mu,\nu,\rho,\sigma $
	
	\item[$ P_D^n $:] 
	$ R_{0z0z} = \frac{4GP}{c^5} \frac{D-z}{r_D^3} , \quad 
	R_{0z0i}=0, \quad R_{0i0j} = \frac{4GP}{c^5} \frac{(D-z)}{r_D\rho^2}\Big(\delta_{ij}-\frac{r_i r_j}{r_D^2\rho^2}(2r_D^2+\rho^2)\Big), \quad
	R_{0izj} = \frac{4GP}{c^5}\frac{1}{\rho^2}\left(\delta_{ij}-\frac{2 r_i r_j}{\rho^2}\right) $

	\item[$ \triangleleft^n $:] $ R_{\mu\nu\rho\sigma} = 0 \quad\forall \mu,\nu,\rho,\sigma. $
\end{itemize}
In the limit $ \rho \ll r,r_D $, the only independent components of the curvature in leading order are
\begin{align}\label{eq:R_oscill_pulse}
R_{0i0j} =&- R_{0izj}= \frac{4GP}{c^5} \frac{1}{\rho^2}\Big(\delta_{ij} - 2 \frac{r_i r_j}{\rho^2}\Big) \equiv \mathcal{R} \quad& \forall (z,t) \in P_0^n\;
\\
R_{0i0j} =& \mathcal{R} = R_{0izj} \quad& \forall (z,t) \in P_D^n,
\end{align}
with $ \rho \ll z,z-D $.
A simplified metric perturbation resulting in the same curvature tensor as equation \eqref{eq:R_oscill_pulse} is given by
\begin{equation}\label{eq:hsimple}
	\tilde{h}^+ = \begin{cases}
		-\frac{8GP}{c^5} \ln \rho & \text{ for } (z,t)\in P_0^n
		\\
		0 & \text{ else }
	\end{cases},
	\quad
	\tilde{h}^- = \begin{cases}
	-\frac{8GP}{c^5} \ln \rho & \text{ for } (z,t)\in P_D^n
	\\
	0 & \text{ else }
	\end{cases}.
\end{equation}
The geodesic equation for a test particle at position $ x^\mu $ is given in coordinate time $ t = \frac{1}{c} x^0 $ by
\begin{equation}
	\frac{{\rm d}^2 x^\mu}{{\rm d}t^2} = - \Gamma^\mu_{\; \alpha\beta} \frac{{\rm d}x^\alpha}{{\rm d}t}\frac{{\rm d}x^\beta}{{\rm d}t} +  \Gamma^0_{\; \alpha\beta} \frac{{\rm d}x^\alpha}{{\rm d}t}\frac{{\rm d}x^\beta}{{\rm d}t}\frac{{\rm d}x^\mu}{{\rm d}t}
\end{equation}
For a non-relativistic test particle this reduces to
\begin{equation}
	\ddot{x}^a = - c^2 \Gamma^a_{\;00} + \mathcal{O}\left(\frac{v^2}{c^2}\right),
\end{equation}
with the linearized Christoffel symbol
\begin{equation}
	\Gamma^\rho_{\;\mu\nu} = \frac{1}{2} \eta^{\lambda \rho}(\partial_\mu h_{\nu\lambda} + \partial_\nu h_{\lambda \mu} - \partial_\lambda h_{\mu\nu}) 
	\implies
	\Gamma^a_{\;00} = -\frac{1}{2} \partial_a \tilde{h}_{00}.
\end{equation}
The acceleration a non-relativistic sensor experiences is therefore equivalent to that from a Newtonian potential
\begin{equation}\label{eq:newtonian_potential}
\varPhi =  \frac{4GP}{c^3}\ln \rho
\end{equation}
for the duration of the pulse passing by ($ P_0,P_D $) with the potential vanishing at all other times.

\section{Intensity in a Fabry-Pérot resonator} \label{app:c}
The considerations here follow those from \cite{cesini_1977} closely but are modified to reflect the setups used in this work.

For a Fabry-Pérot resonator consisting of two mirrors with field reflection coefficients $ \sqrt{R_1},\sqrt{R_2} $ and field transmission coefficients $ \sqrt{T_1},\sqrt{T_2} $, the field in cavity (at the face  of mirror 1) resulting from a pump beam striking mirror 1 can be written as
\begin{equation}
	E_{\rm cav} (t) = \sqrt{R_1R_2} E_{\rm cav}(t-\tau_{\rm rt}) + \sqrt{T_1}E_{\rm pump}(t)
\end{equation}
in the time domain, where $ \tau_{\rm rt} $ is the time for one round trip in the cavity.
In the frequency domain this can be written as
\begin{equation}\label{eq:cavity_fourieramp}
	\tilde{E}_{\rm cav} (\omega) = \tilde{G}(\omega) \tilde{E}_{\rm pump}(\omega), \text{ with } \tilde{G}(\omega) = \frac{\sqrt{T_1}}{1-\sqrt{R_1R_2}e^{-i\omega\tau_{\rm rt}}}.
\end{equation}
\subsection{Single monochromatic rectangular pulse}
For a monochromatic pump field of frequency $ \omega_{\rm E} $ and length $ \tau_{\rm p} $ entering the cavity at $ t=0 $ the pump field is given by
\begin{align}\label{eq:pump_field_single_pulse}
	E^{\rm p}_{\rm pump}(t) &= E_0 e^{i\omega_{\rm E} t} \textifsym{l|H|l}_{\tau_{\rm p}}(t), \text{ with  }\textifsym{l|H|l}_{\tau_{\rm p}}(t) =\Theta\left(t\right) - \Theta\left(t-\tau_{\rm p}\right)
	\\
	\implies \tilde{E}^{\rm p}_{\rm pump} (\omega) &= E_0 \tau_{\rm p} e^{-i\omega \tau_{\rm p}/2} {\rm sinc}( (\omega-\omega_{\rm E})\tau_{\rm p}).
\end{align}
The corresponding Fourier transformed field amplitude in the cavity is then given through eq.\eqref{eq:cavity_fourieramp} by
\begin{align}\label{eq:Ecav_pulse}
	\tilde{E}^{\rm p}_{\rm cav} (\omega) &= E_0 \frac{\sqrt{T_1}e^{-i\omega\tau_{\rm p}/2}}{1-\sqrt{R_1R_2}e^{-i\omega\tau_{\rm rt}}} \tau_{\rm p}{\rm sinc} ((\omega-\omega_{\rm E})\tau_{\rm p})
	\\
	\implies E^{\rm p}_{\rm cav} (t) &= E_0 \sqrt{T_1} \sum_{n=0}^{\infty} \left(R_1R_2\right)^{n/2} e^{i\omega_{\rm E}(t-n \tau_{\rm rt})} \textifsym{l|H|l}_{\tau_{\rm p}}(t-n\tau_{\rm rt})
\end{align}
For a very short pulse $ \tau_{\rm p}\ll \tau_{\rm rt} $, none of the addends will overlap and the intensity in the cavity is
\begin{equation}
	I^{\rm p}_{\rm cav} (t) = \left|E^{\rm p}_{\rm cav} (t)\right|^2 = E_0^2 T_1 \sum_{n=0}^\infty (R_1R_2)^n \textifsym{l|H|l}_{\tau_{\rm p}}(t-n\tau_{\rm rt}).
\end{equation}
The pulse enters the cavity with an intensity reduced by $ T_1 $ and is reduced by a further factor $ R_1R_2 $ for each subsequent round trip.
The average power in the cavity relative to that of the pump laser is then given by 
\begin{equation}\label{eq:cavityenhancement}
	\frac{P_{\rm cav}}{P_0} = T_1 \sum_{n=0}^\infty (R_1R_2)^n = \frac{T_1}{1 - R_1R_2} =\frac{1-R_1}{1-R_1R_2}\lesssim 1.
\end{equation}

\begin{figure}
	\centering
	\includegraphics[width=0.7\linewidth]{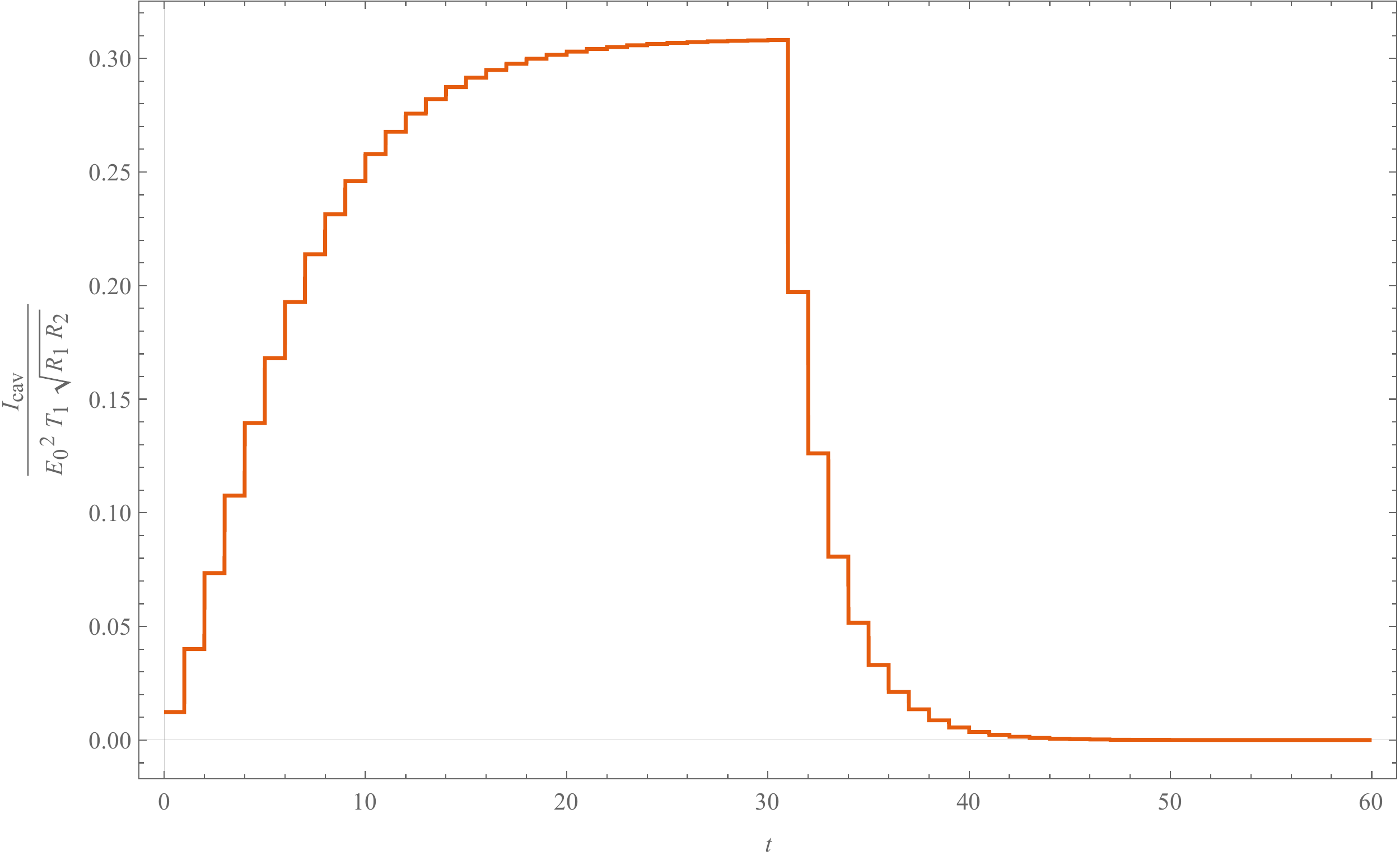}
	\caption{Intensity inside the cavity resulting from a long ($ \tau_{\rm p} \gg \tau_{\rm rt} $) rectangular pulse of monochromatic or spectrally narrow light. The time is given in units of $\tau_{\rm rt}$.}
\label{fig:jaggedsharkfin}
\end{figure}

For long pulses ($ \tau_{\rm p} \gg \tau_{\rm rt} $) and a resonant cavity ($ \omega_{\rm E}\tau_{\rm rt} = 2\pi m $) only addends from $ n_{\rm min} = \max\left( 0,\left\lceil \frac{t-\tau_{\rm p}}{\tau_{\rm rt}}\right\rceil\right) $ to $ n_{\rm max} = \max\left( 0,\left\lfloor \frac{t}{\tau_{\rm rt}} \right\rfloor \right)$ contribute for any given time, returning
\begin{align}
	E^{\rm p}_{\rm cav} (t) =& E_0 \sqrt{T_1} \left[\frac{1-\left(R_1R_2\right)^\frac{n_{\rm max}+1}{2}}{1+\sqrt{R_1R_2}}-\frac{1-\left(R_1R_2\right)^\frac{n_{\rm min}+1}{2}}{1+\sqrt{R_1R_2}}\right] 
	\\
	=&  E_0 \sqrt{T_1} \sqrt{R_1R_2} \frac{\left(R_1R_2\right)^{n_{\rm min}/2}-\left(R_1R_2\right)^{n_{\rm max}/2}}{1+\sqrt{R_1R_2}}.
\end{align}
The intensity for this long-pulse resonant cavity case can be described as a ``jagged shark fin'' and is plotted in fig. \ref{fig:jaggedsharkfin}

\subsection{Series of monochromatic rectangular pulses}
A series of periodic pulses separated by time $ \tau_{\rm rt} $ can be written as a sum of pulses $  E^\Sigma_{\rm pump} (t) = \sum_k E^{\rm p}_{\rm pump} (t-k\tau_{\rm rep}) $. As all of the operations on the field are linear the pump field eq.~\eqref{eq:Ecav_pulse} can be used to find
\begin{equation}
	E^\Sigma_{\rm cav} (t) = \sum_k E^{\rm p}_{\rm cav} (t-k\tau_{\rm rep}).
\end{equation}
For repetition times much 
longer than the lifetime of a pulse in the cavity $ \tau_{\rm rep} \gg\tau_{\rm L} \sim \frac{\tau_{\rm rt} (R_1R_2)^{1/4}}{1-\sqrt{R_1R_2}}  $, and the pulse length $ \tau_{\rm rep} \gg\tau_{\rm p} $, the addends barely overlap such that there is no interference between consecutive pulses. In this case the intensity in the cavity is just that of the singular pulse repeating periodically.
\bibliography{mybibs_bt}
\end{document}

%% file: spacetimezones.pdf_tex
\begingroup%
  \makeatletter%
  \providecommand\color[2][]{%
    \errmessage{(Inkscape) Color is used for the text in Inkscape, but the package 'color.sty' is not loaded}%
    \renewcommand\color[2][]{}%
  }%
  \providecommand\transparent[1]{%
    \errmessage{(Inkscape) Transparency is used (non-zero) for the text in Inkscape, but the package 'transparent.sty' is not loaded}%
    \renewcommand\transparent[1]{}%
  }%
  \providecommand\rotatebox[2]{#2}%
  \newcommand*\fsize{\dimexpr\f@size pt\relax}%
  \newcommand*\lineheight[1]{\fontsize{\fsize}{#1\fsize}\selectfont}%
  \ifx\svgwidth\undefined%
    \setlength{\unitlength}{176.32922539bp}%
    \ifx\svgscale\undefined%
      \relax%
    \else%
      \setlength{\unitlength}{\unitlength * \real{\svgscale}}%
    \fi%
  \else%
    \setlength{\unitlength}{\svgwidth}%
  \fi%
  \global\let\svgwidth\undefined%
  \global\let\svgscale\undefined%
  \makeatother%
  \begin{picture}(1,1.50954791)%
    \lineheight{1}%
    \setlength\tabcolsep{0pt}%
    \put(0,0){\includegraphics[width=\unitlength,page=1]{spacetimezones.pdf}}%
    \put(0.73252731,0.01000215){\color[rgb]{0,0,0}\makebox(0,0)[lt]{\lineheight{1.25}\smash{\begin{tabular}[t]{l}$D$\end{tabular}}}}%
    \put(0.13148907,0.00582608){\color[rgb]{0,0,0}\makebox(0,0)[lt]{\lineheight{1.25}\smash{\begin{tabular}[t]{l}0\end{tabular}}}}%
    \put(0.03192845,1.45784243){\color[rgb]{0,0,0}\makebox(0,0)[lt]{\lineheight{1.25}\smash{\begin{tabular}[t]{l}$ct$\end{tabular}}}}%
    \put(0.88333037,0.0151404){\color[rgb]{0,0,0}\makebox(0,0)[lt]{\lineheight{1.25}\smash{\begin{tabular}[t]{l}$z$\end{tabular}}}}%
    \put(0,0){\includegraphics[width=\unitlength,page=2]{spacetimezones.pdf}}%
    \put(0.48459617,0.73810774){\makebox(0,0)[lt]{\lineheight{1.25}\smash{\begin{tabular}[t]{l}$\triangleleft$\end{tabular}}}}%
    \put(0.29569832,0.22564288){\makebox(0,0)[lt]{\lineheight{1.25}\smash{\begin{tabular}[t]{l}$\triangleright$\end{tabular}}}}%
    \put(0.41162642,1.04158313){\makebox(0,0)[lt]{\lineheight{1.25}\smash{\begin{tabular}[t]{l}$P_0$\end{tabular}}}}%
    \put(0.40800953,0.4613983){\makebox(0,0)[lt]{\lineheight{1.25}\smash{\begin{tabular}[t]{l}$P_{\rm D}$\end{tabular}}}}%
  \end{picture}%
\endgroup%